%% file: main.tex
\documentclass[sigconf,natbib=true,anonymous=false]{acmart}

\AtBeginDocument{%
  \providecommand\BibTeX{{%
    \normalfont B\kern-0.5em{\scshape i\kern-0.25em b}\kern-0.8em\TeX}}}

\setcopyright{acmcopyright}
\copyrightyear{2021}
\acmYear{2021}
\acmDOI{10.1145/1122445.1122456}

\acmConference[Woodstock '18]{Woodstock '18: ACM Symposium on Neural
  Gaze Detection}{June 03--05, 2018}{Woodstock, NY}
\acmBooktitle{Woodstock '18: ACM Symposium on Neural Gaze Detection,
  June 03--05, 2018, Woodstock, NY}
\acmPrice{15.00}
\acmISBN{978-1-4503-XXXX-X/18/06}

\usepackage{booktabs}
\usepackage{multirow}
\usepackage{graphicx}
\usepackage{stfloats}
\usepackage{comment}
\usepackage{pgf}
\usepackage{graphicx}

\begin{document}

\title{On the Overlooked Significance of Underutilized Contextual Features in Recent News Recommendation Models}


\author[ ]{Sungmin Cho}
\affiliation{
  \institution{Seoul National University}
  \country{South Korea}
}
\email{tjdals4565@gmail.com}

\author[ ]{Hongjun Lim}
\affiliation{
  \institution{NAVER Corp}
  \country{South Korea}
}
\email{hongjun.lim@navercorp.com}

\author[ ]{Keunchan Park}
\affiliation{
  \institution{NAVER Corp}
  \country{South Korea}
}
\email{keunchan.park@navercorp.com}

\author[ ]{Sungjoo Yoo}
\affiliation{
  \institution{Seoul National University}
  \country{South Korea}
}
\email{sungjoo.yoo@gmail.com}

\author[ ]{Eunhyeok Park}
\affiliation{
  \institution{POSTECH}
  \country{Seoul, South Korea}
}
\email{canusglow@gmail.com}

\renewcommand{\shortauthors}{Short Authors}

\begin{abstract}
\input{sections/abstract}

\end{abstract}





\maketitle
\pagestyle{plain}

\section{Introduction}
\input{sections/intro}
\section{Related Work}
\input{sections/relworks}

\section{Feature Definitions}
\input{sections/feature_definitions}

\section{Reevaluating the Importance of contextual features}
\input{sections/boosting}

\section{Deep Learning Solution}
\input{sections/deeplearning}

\section{Conclusion \& Future works}
\input{sections/conclusion}

\begin{acks}
Acknowledgements.
\end{acks}

\bibliographystyle{ACM-Reference-Format}
\bibliography{bibs}


\end{document}

%% file: sections/abstract.tex

Personalized news recommendation aims to provide attractive articles for readers by predicting their likelihood of clicking on a certain article. To accurately predict this probability, plenty of studies have been proposed that actively utilize content features of articles, such as words, categories, or entities. However, we observed that the articles' contextual features, such as CTR (click-through-rate), popularity, or freshness, were either neglected or underutilized recently. To prove that this is the case, we conducted an extensive comparison between recent deep-learning models and naive contextual models that we devised and surprisingly discovered that the latter easily outperforms the former. Furthermore, our analysis showed that the recent tendency to apply overly sophisticated deep-learning operations to contextual features was actually hindering the recommendation performance. From this knowledge, we design a purposefully simple contextual module that can boost the previous news recommendation models by a large margin.

%% file: sections/intro.tex
Nowadays, the majority of news consumption happens online. Online news platforms seek to assemble news articles from multiple sources and provide a personalized recommendation of these articles for each user. These services face unique challenges since the lifespan of each news is very short, the volume of articles and the number of users are very large, and users have very diverse and dynamic interests. Therefore, the system requires a news recommendation model that can effectively model the relevance of news articles in a timely manner~\cite{das2007google, liu2010personalized, phelan2009using, phelan2011terms, son2013location}.

News recommendation models have benefited from the recent advances in deep learning methods in terms of news modeling and user modeling. Since deep learning methods are capable of creating meaningful representations in embedding space, they were used to create succinct news representations from the articles' content features such as words, images, and categories~\cite{DSSM, Okura, Weave, DKNwang2018a, NRMS, NAML, NPA, wu2021empowering}. Likewise, deep learning techniques were also used to capture the preferences of users by generating user representations from their past interactions with the system~\cite{song2016, Okura, kumar2017, HRAM, an2019, zhu2019, zhang2019a, park2017deep, Weave, FIM, NRMS, NAML, NPA, DKNwang2018a, wu2021feedrec, MMRec, wu2021empowering}. These representations were then used to measure the relevance of an article to a user, which ultimately led to better recommendation performance.

While the research progress has been immense in modeling these content features, we claim that contextual features (for instance, the trendiness of an article) haven't been paid the same amount of attention recently. To show this, we relied on a recent thorough survey~\cite{wuchuhansurvey} to gather the published papers on these subjects, then plotted their distributions over the years in Figure~\ref{fig:content_context}. As it shows, the amount of research on utilizing contextual features of news articles has continuously been fairly low compared to the amount of research on utilizing content features which has been rapidly increasing with the advent of deep learning. However, considering that news consumption often happens with regard to its context (e.g., how "hot" is this subject right now?), there seems to be a need to reevaluate the importance of the contextual features and their potential for improving recommendation quality.

In this paper, we attempt such reevaluation as follows. First, we select and define the contextual features that prove to be consistently effective across many datasets. Then, we devise a number of naive contextual models that work with these features. By showing that these naive models easily outperform recent content-based deep-learning models, we show the overlooked importance of the contextual features, and analyze the reasons for such underutilization. From this analysis, we finally suggest a deep learning module that can boost the accuracy of previous content-based models by a large margin. We present extensive experiments on four real-world datasets where we achieve state-of-the-art performance.

%% file: sections/relworks.tex
The goal of this section is mainly threefold. First, in section \ref{newsrecommendationsubsection}, we aim to summarize recent deep learning based approaches to news recommendation. We only covered deep learning models here since there is a big performance gap between them and the more traditional methods (e.g. Bag-of-Words, TF-IDF, etc)\footnote{For a more thorough history we refer the reader to \cite{wuchuhansurvey}.}. These works naturally consist of content-based models because of their abundance in recent publications. Second, in section \ref{contextualsubsection}, we trace the history of works that tried to utilize contextual features of news articles. Unlike section \ref{newsrecommendationsubsection}, in section \ref{contextualsubsection} we also include traditional works to give our work the right context. Few deep-learning models that are included here are excluded from section \ref{newsrecommendationsubsection}. Finally, in section \ref{boostingsubsection}, we briefly touch on gradient boosting methods that we used in section \ref{section4}.

\subsection{News Recommendation}
\input{sections/relworks_subsections/news}

\subsection{Contextual Features}
\input{sections/relworks_subsections/contextual}

\subsection{Gradient Boosting}
\input{sections/relworks_subsections/boosting}




%% file: sections/relworks_subsections/news.tex
\label{newsrecommendationsubsection}
\begin{figure}[t]
  \centering
  \includegraphics[width=1.0\linewidth]{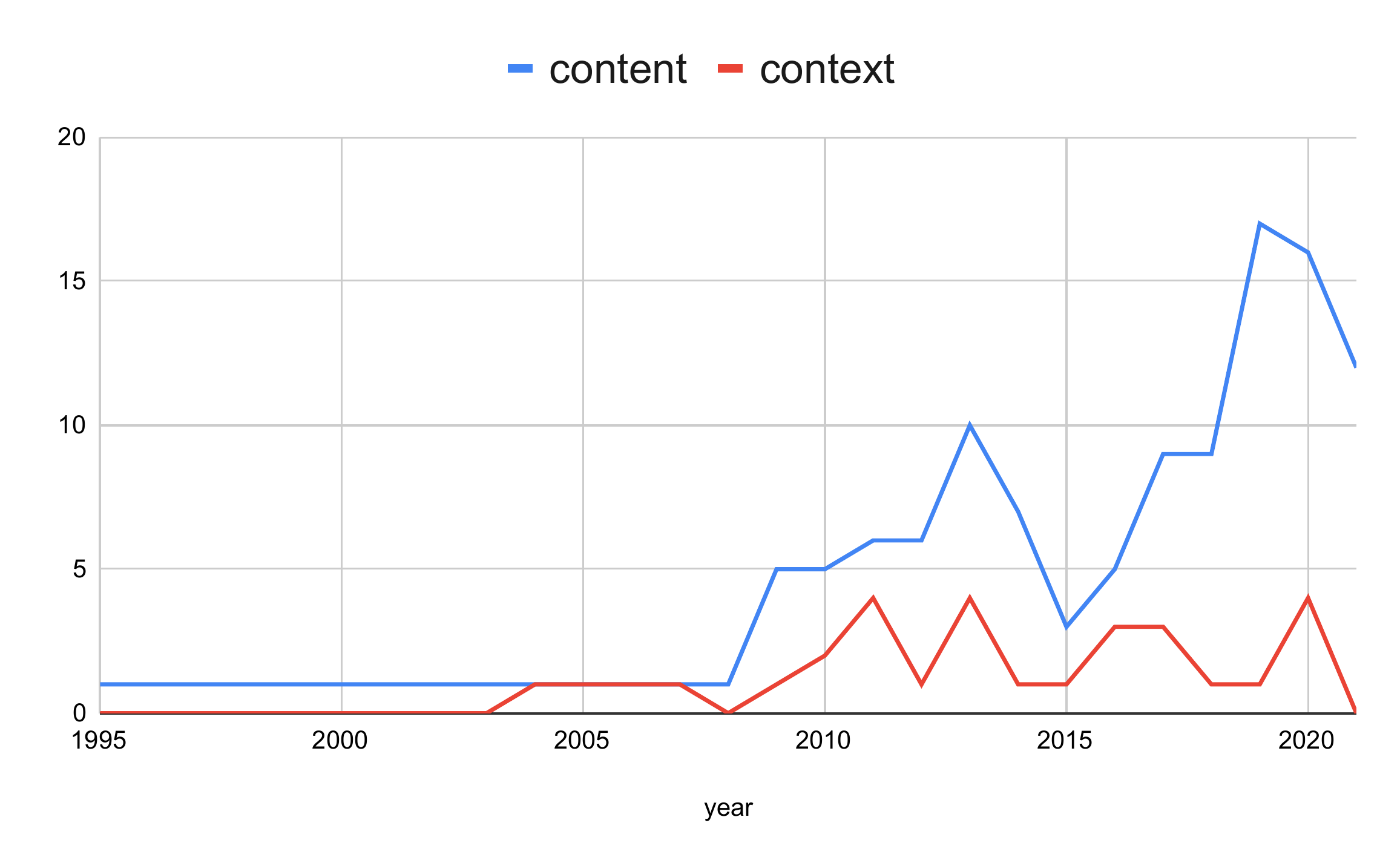}
  \caption{Number of news recommendation papers on each subject by their publication years (reconstructed from News Modeling section in \cite{wuchuhansurvey}). Content features include: words, topics, entities, categories, etc. Contextual features include: popularity, recency, novelty, location, etc.}
  \Description[Test Desc]{Test Description}
  \label{fig:content_context}
\end{figure}

\begin{figure}[t]
  \centering
  \includegraphics[width=0.8\linewidth]{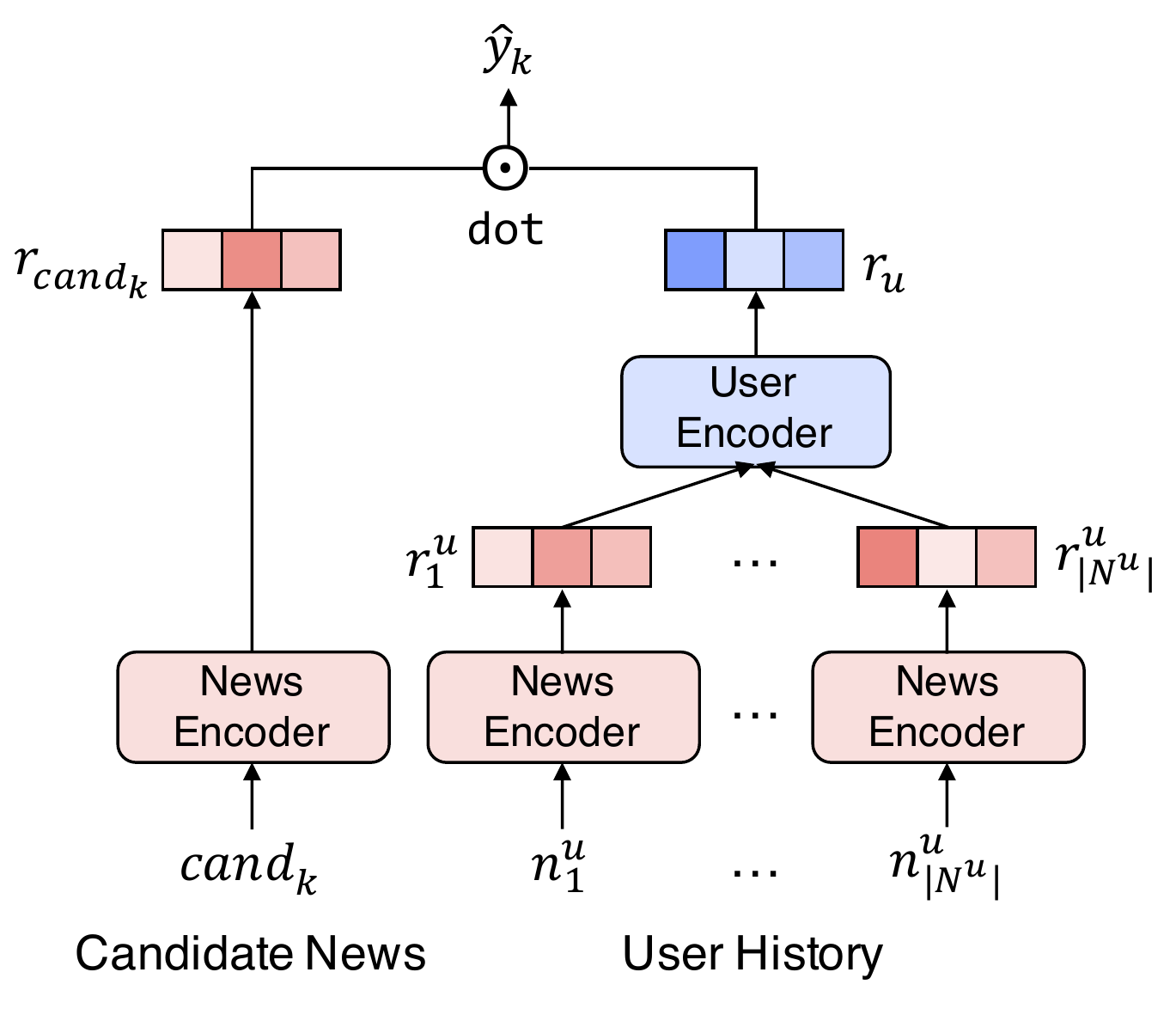}
  \caption{A typical framework for news recommendation where the user representation is compared with candidate news representations by a dot product operation.}
  \Description[Test Desc]{Test Description}
  \label{fig:typical}
\end{figure}



\subsubsection{Problem Formulation}
The problem of news recommendation can be formulated as follows. Let $\mathcal{U}$ be a set of users and $\mathcal{N}$ be a set of news articles. The history of each user $u \in \mathcal{U}$ can be represented as $\mathcal{N}^u=[(n^u_1, t^u_1), ..., (n^u_{|\mathcal{N}^u|}, t^u_{|\mathcal{N}^u|})]$ where $n^u_k \in \mathcal{N}$ is the $k$th news article clicked by $u$ at timestamp $t^u_k$. 

Given the above, our goal is to recommend a set of news articles for a user at target timestamp. More specifically, at target time $t_{target}$, given $K$ candidate news articles $[cand_1, ..., cand_{K}]$ and the user's history $[n^u_1, ..., n^u_{|N^u|}]$, our goal is to rank those candidates in the order of the user's estimated preferences. 

\subsubsection{Architecture}
Figure \ref{fig:typical}. illustrates a typical architecture for news recommendation~\cite{NRMS}, which roughly captures most recent deep-learning based news recommendation models. Their approach is similar in the sense that they compare the user vector with the vectors of candidate news articles based on similarity-measure operations such as dot product. Therefore, central to the whole framework are two components that create these vectors: news encoder and user encoder. A typical news encoder~\cite{NRMS} encodes a news article into a low-dimensional vector. Then a typical user encoder~\cite{NRMS} generates a user representation by aggregating news representations from the user's history that are each encoded by the former news encoder. In this section, we will describe in detail how previous works approached each one of these components.

\subsubsection{News Modeling}
As news is inherently textual, many works focused on generating news representations from the text.
DSSM~\cite{DSSM} was an early attempt to learn representations of web documents and queries by using letter n-gram based word hashing and dense layers to project them into low-dimensional vectors.
Okura et al.~\cite{Okura} proposed to apply denoising autoencoder to news articles to get distributed representations.
Meanwhile, other works proposed to use Convolutional Neural Networks (CNNs) in various ways~\cite{Weave, NAML, NPA, an2019, DKNwang2018a, zhu2019, zhang2019a, FIM}. For example, Weave\&Rec~\cite{Weave} applied CNN on word2vec representations of words in news articles. DKN~\cite{DKNwang2018a} designed a KCNN module to make use of knowledge entities aligned with news titles.
More recently, attention mechanisms were used to boost the power of news encoders~\cite{NAML, NRMS, NPA, ge2020graph, wu2021feedrec, MMRec, wu2021empowering}. For example, NRMS~\cite{NRMS} proposed to apply self-attention and additive attention to the word embeddings to generate news representations that are aware of the importance of each word. NAML~\cite{NAML} suggested to include (sub)category information of news articles by using (sub)category embeddings together with word representations in an attention operation. Wu et al.~\cite{wu2021empowering} showed that one can empower the news encoder by using pre-trained language models (e.g. BERT~\cite{BERT}), which seems to imply that the power of news encoders will continue to improve as long as there are advances in NLP research.

\subsubsection{User Modeling}
Deep learning methods were shown to be much more effective than traditional methods in terms of user modeling. For example, Okura et al. \cite{Okura} reported significant performance improvement over traditional temporal methods such as time decay when they used Recurrent Neural Networks (RNNs) to model user behaviors. In fact, since GRU4Rec~\cite{GRU4Rec} was first proposed, many sequential recommendation methods employed RNN to benefit from its sequence understanding capability~\cite{hidasi2018recurrent, RNN4Rec, HIERARCHICALGRU4REC, DREAM, wu2017recurrent, wu2017sequential, DINEVOLUTION}. The same trend occurred in the news domain for user modeling~\cite{song2016, Okura, kumar2017, HRAM, an2019, zhu2019, zhang2019a, park2017deep}. For example, LSTUR~\cite{an2019} proposed to use GRU on the users' history to learn their short-term representations while their long-term representations were learned by ID-based embeddings.
Some found Convolutional Neural Networks (CNNs) to be useful~\cite{Weave, FIM}. For example, Weave\&Rec~\cite{Weave} applied 3D CNN to the user's behavior sequence where each item was represented as a 2D matrix comprised of word embeddings.
More recently, prevalent usage of attention mechanisms were witnessed in recommendation models. In sequential recommendation, following the huge success of transformers in NLP~\cite{TRANSFORMER, BERT, TRANSFORMER-XL, XLNET}, models such as SASRec~\cite{SASRec}, BERT4Rec~\cite{BERT4Rec} and other variants~\cite{TISASRec} 
fully adopted attention techniques, in contrast to previous models where attention was merely a supplementary part of the model~\cite{NARM, STAMP}. The same was true for user modeling in news domain~\cite{NRMS, NAML, NPA, DKNwang2018a, zhu2019, zhang2019a, wu2021feedrec, MMRec, wu2021empowering}. For example, models such as NRMS~\cite{NRMS}, NAML~\cite{NAML} and NPA~\cite{NPA} all used similar attention operations to combine the user's history into a single representation.

%% file: sections/relworks_subsections/contextual.tex
\label{contextualsubsection}
In contrast to content features which capture relatively static information about news articles and users, contextual features try to capture their more dynamic aspects. For example, popularity of an article is an important signal that measures its relevance to the mass at a certain point of time~\cite{SCENE, dynamicbilinear, penrecsys, newspersonalizationusingsupportvectormachines, personalizednewsrecommendationbasedonimplicitfeedback, twitterpopularity, facebooktwitterlocation, hypner, logo, liu2016research, tailored, incorporatingpopularity, utilitybased}. For the same reason, CTR (click-through-rate) of an article is also an important signal since it measures the article's general attractiveness ~\cite{dynamicbilinear, pprec}.  Freshness/Recency should also be taken into account to consider the staleness of an article ~\cite{newspersonalizationusingsupportvectormachines, facebooktwitterlocation, SCENE, hypner, logo, pprec, tailored, utilitybased}. Some works incorporate other contexts such as location \cite{tailored, personalizednewsrecommendationbasedonimplicitfeedback, facebooktwitterlocation, son2013location, viana2017hybrid, yeung2010proactive} or weather \cite{yeung2010proactive}.

Central difference between these works lies in the way they choose to encode these signals. For example, \cite{hypner} suggested to use number of views of an article as its popularity, whereas \cite{SCENE} suggested to encode the article popularity as $\frac{\text{\# clicked users}}{\text{\# all users}}$, and \cite{newspersonalizationusingsupportvectormachines} suggested $log(1 + \text{\# clicked users})$ to represent it\footnote{\# meaning "number of"}. 
An article's recency was represented as $\frac{\Delta t}{24*60}$ in \cite{SCENE}, while $log(1+\Delta t)$ was used in \cite{newspersonalizationusingsupportvectormachines} \footnote{$\Delta t$ = current time - article publish time}.
For CTR, \cite{dynamicbilinear} proposed to estimate the CTR at event time by using Kalman filters, whereas \cite{pprec} chose to use past CTR either as its raw value (in their popularity predictor module), or by quantifying its value then doing an embedding look-up (in their popularity-aware user encoder).

From this we would like to note two things. First, judging by the number of publications, some features such as past CTR haven't been paid much attention despite being very important signals. In fact, experiments in \cite{pprec} suggest that a simple baseline model that rank articles only by their CTRs can match the performance of heavy deep learning models. Our work seeks to further explore this discovery.
Second, the encoding methods for these features remained rather primitive, and the effect of recent non-linear methods such as deep learning on these features needs to be reevaluated, which is precisely what we aim in this work.

%% file: sections/relworks_subsections/boosting.tex
\label{boostingsubsection}

Gradient boosting method is a machine technique that trains a prediction model from an ensemble of decision trees. It grows decision trees by progressively reducing the residual errors from the previous trees. Gradient boosting libraries such as XGBoost \cite{xgboost}, LightGBM \cite{lightgbm}, and CatBoost \cite{catboost} are currently go-to solutions for many data science and machine learning challenges that deal with tabular data. In fact, although the prominent approach for recommendation tasks have shifted to deep learning, gradient boosting methods persistently appear as winner solutions in recommendation challenges \cite{recsys2019challengewinner, recsys2020challengewinner}. Also, a recent paper \cite{hyperconnectboosting} surprisingly discovered that gradient boosting methods can be both more efficient and more effective than over-parameterized deep learning methods in CTR prediction task.




%% file: sections/feature_definitions.tex
\label{feature_definitions}

In this section we define the contextual features that will be used throughout the rest of the paper.

\subsection{CTR}
Click-through-rate of an article is naturally defined as follows:

\begin{equation}
    CTR = \frac{\text{number of clicks}}{\text{number of impressions}}
\end{equation}

In other words, it is a ratio of the number of clicks to the total number of times it has been shown to all users. Since higher CTR usually indicates successful recommendation, predicting CTRs has been a popular optimization task for recommendation models~\cite{hyperconnectboosting}. However, one should note that it can also be used as a feature of an item that describes its attractiveness, and therefore as an input to the recommendation model. In an offline evaluation setting, one should be careful not to incorporate current and future records when creating features since it causes data leakage problem. Therefore, to use as a valid feature, we parameterize CTR with current time $t$:

\begin{equation}
    CTR(t) = \frac{\text{number of clicks before $t$}}{\text{number of impressions before $t$}}
\end{equation}

\subsection{Popularity}
We use two popularity features. First is the obvious number of clicks:
\begin{equation}
    numclicks(t) = \text{number of clicks before $t$}
\end{equation}

The downside of this feature is that it is not affected by the temporal distance between now and the event of the click. However, in reality, even articles that have a large amount of clicks should not be considered as being popular if the clicks happened a long time ago. To incorporate this temporal influence, we suggest a second popularity feature called \textit{trendiness}:

\begin{equation}
    trendiness(t) = \sum_{k}{m(t, k)}
\end{equation}

\begin{equation}
    m(t, k) =
\begin{cases}
    e^{-\alpha (t - t_k)},& \textrm{if } t \geq\ t_k \label{eqn:alpha} \\ 
    0,              & \textrm{otherwise}
\end{cases}
\end{equation}

where $m(t,k)$ is the "mass" of the $k^{th}$ click of the article at time $t$, $t_k$ is the timestamp of the $k^{th}$ click, and $\alpha$ is a hyperparameter to control the degree of decay (we use $\alpha=0.001$ in this paper). In other words, every click generates a mass of 1 at the time of click which gets exponentially decayed as time passes. The article's trendiness is defined as the accumulation of these decayed masses.
The benefit of this feature is that it captures both the popularity and its temporal relevance, and that it is computationally simple to maintain its value because exponential functions support easy updates (multiplications can be chained).

\subsection{Freshness}
Following previous works, we define freshness feature as:

\begin{equation}
    freshness(t) = t - \text{publish time}
\end{equation}

In other words, it measure how much time has passed since the publication of the article. We use seconds as the unit of time.

\subsection{Discussion}
Although we settle for these four features in this paper, note that the list is not exhaustive and that more features can be imagined. We have tried other features such as CTR in a particular user group (e.g. groups by age and gender), CTR in a particular time window (e.g. pertaining only to the last 10 minutes), timestamp (or hour) of the event of recommendation, etc. However, we omitted them since they weren't proved to be consistently effective across all datasets. Engineering more useful features remains as a future work.

%% file: sections/boosting.tex
\label{section4}
This section aims to prove the overlooked importance of contextual features with the following strategy. We prepare four real-word news recommendation datasets for solid comparison between models.
Then, on the other hand, we prepare recent deep-learning news recommendation models that heavily make use of content features of articles, some of which tries to also incorporate contextual features. In addition, we devise naive contextual models that make use of contextual features in a purposefully naive way. By showing that the latter models easily achieve superior performance, we emphasize the overlooked significance of contextual features and assess the reasons for such underutilization in the past.

\subsection{Datasets}
\input{sections/boosting_subsections/datasets}

\subsection{Evaluation}
\input{sections/boosting_subsections/evaluation}

\input{sections/boosting_subsections/dataset_details_table}

\input{sections/boosting_subsections/five_models_figure}

\subsection{Methods}
\input{sections/boosting_subsections/methods}

\input{sections/boosting_subsections/results_table}
\input{sections/boosting_subsections/pprec_ablation_table}

\subsection{Performance Comparison}
\input{sections/boosting_subsections/results}

%% file: sections/boosting_subsections/datasets.tex
We used three public datasets (\textbf{Mind} \cite{minddataset}, \textbf{Globo} \cite{globodataset}, and \textbf{Adressa} \cite{adressadataset}) and one proprietary dataset (\textbf{Prop}) to conduct experiments.
\textbf{Mind} is a large English news recommendation dataset used for a recent challenge \cite{minddataset} collected from Microsoft News \footnote{https://microsoftnews.msn.com/}
\footnote{Mind dataset has two versions: MIND and MIND-SMALL. We used MIND-SMALL in this paper for faster experiments.}. 
\textbf{Adressa} \footnote{Adressa dataset has two versions: 1w (one week) and 10w (ten weeks). We used 1w for this paper.} is a large Norwegian news recommendation dataset collected from Adresseavisen \footnote{https://www.adressa.no/}.
\textbf{Prop} is a large news recommendation dataset collected from a large news service company \footnote{to be revealed after the publication of this paper}.
The details of these datasets can be found in Table \ref{tab:dataset_details}.
As the table shows, some datasets lack important information. Since \textbf{Globo} and \textbf{Adressa} don't have impressions, CTR can't be computed for them. Since \textbf{Mind} lacks publish timestamps of news articles, freshness can't be computed. Models that utilize entity \footnote{Entity denotes recognizable objects in news content, for example a famous person or an event.} embeddings can only be applied to \textbf{Mind}.
As a temporary solution, features which can't be computed were omitted from the experiments for that dataset.

%% file: sections/boosting_subsections/evaluation.tex
We sorted the targets by their timestamps and split them by 8:1:1 to get train, validation, and test set respectively. For datasets with impressions (\textbf{Mind} and \textbf{Prop}), the unclicked items in the same impression were used as negative items. $K=4$ items were sampled from them for training, and all negative items were used for evaluation. For datasets without impressions, following \cite{globodataset}, we sampled negative items from the clicks of other users within the recent queue of 10 minutes. $K=4$ and $K=20$ items were sampled for training and evaluation respectively. Model's performance was measured as its capability to rank positive items higher than negative items, which was measured with metrics such as NDCG@k, Recall@k, and AUC. For this section, we sampled 5K users from each dataset and used these subsets for experiments.

%% file: sections/boosting_subsections/dataset_details_table.tex
\begin{table}[t]
\caption{Dataset details. Note that this is after we have filtered and preprocessed the datasets.}
\label{tab:dataset_details}
\small
\centering
\begin{tabular}{@{}l|c|c|c|c@{}}
    \toprule
        \textbf{Dataset} & \textbf{Prop} & \textbf{Mind} & \textbf{Globo} & \textbf{Adressa} \\
    \midrule
        number of users & 87.7K  & 94.1K  & 317.8K & 613.4K \\
        number of items & 64.7K & 52.9K  & 43.7K & 17.4K \\
        number of interactions & 3.1M & 2.4M & 2.8M  & 2.6M \\
        sparsity & 0.055\% & 0.048\% & 0.021\% & 0.025\% \\
        period & 16 days & 7 days & 16 days & 7 days \\
        includes impressions? & \checkmark & \checkmark &  &  \\
        includes publish timestamps? & \checkmark &  & \checkmark & \checkmark \\
        provides text of articles? & \checkmark & \checkmark &  & \checkmark \\
        provides entities? &  & \checkmark &  &  \\
    \bottomrule
\end{tabular}
\end{table}

%% file: sections/boosting_subsections/five_models_figure.tex
\begin{figure*}[t]
  \centering
  \includegraphics[width=1.0\linewidth]{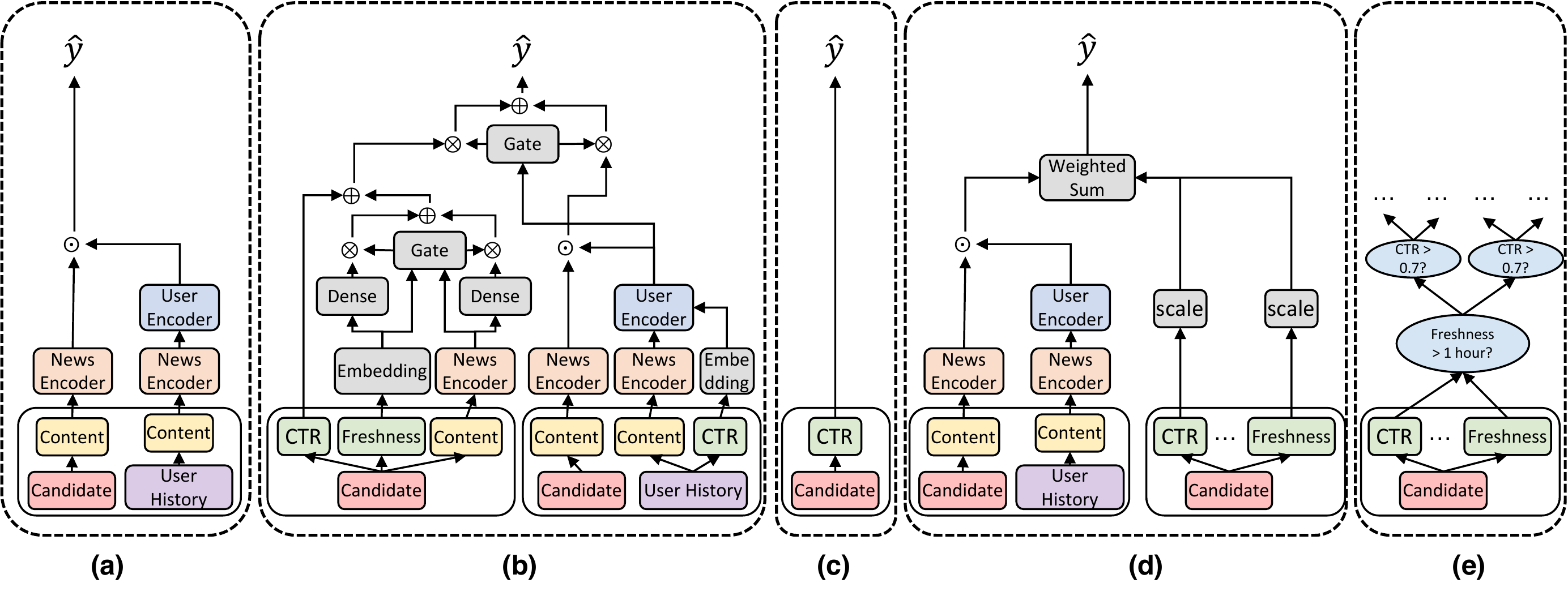}
  \caption{Five kinds of methods we compare. Each method uses various features and operations to predict the probability of the user clicking on the candidate article ($\hat{y}$ in the figure) (a) Previous deep-learning content models (illustrated in detail in Figure \ref{fig:typical}) (b) The architecture of PP-Rec \cite{pprec}. (c) \textbf{Naive-CTR} which uses raw CTR value as the logit. (d) \textbf{NRMS+Naive} which does a weighted sum of content logits and naive contextual logits to make the final prediction. Some contextual features such as $numclicks$ and $trendiness$ are omitted in the figure for the sake of space. The scale operation is identity for CTR and $log(1+x)$ for the rest. (e) Decision tree created by gradient boosting method. The prediction is made by the value of the leaf node.}
  \Description[Test Desc]{Test Description}
  \label{fig:five_models}
\end{figure*}

%% file: sections/boosting_subsections/methods.tex
We compared five different kinds of methods. The first two consist of existing baselines: (1) Previous deep-learning content models and (2) Previous deep-learning content models that incorporated contextual features. Then, to compare against these baselines, we devised three kinds of models: (3) Naive context models, (4) Deep-learning content model with naive context incorporation, and (5) Gradient boosting context models.

First, previous deep-learning content models denote the models described in section \ref{newsrecommendationsubsection}. It is also illustrated in Figure \ref{fig:five_models} (a).
For experiments in this section, we used \textbf{NRMS} to represent this class of models. \textbf{NRMS} is a good representative because many other models can be seen as its variations. For example, if we simply replace its self-attention layers with convolution layers, and additionally use articles' category/subcategory information in its news encoder, we get \textbf{NAML}~\cite{NAML}. Also, it is a good representative because its performance is on par with other models according to \cite{pprec}. For \textbf{Globo}, since it doesn't provide textual information of news articles and instead provides each item's pre-trained embedding, we had to use a variation of \textbf{NRMS} where we replace the output of news encoders with those embeddings.

Second, for previous deep-learning content models that incorporated contextual features, we used \textbf{PP-Rec} \cite{pprec} since there were no other alternatives to our knowledge. Its architecture is illustrated in Figure \ref{fig:five_models} (b).
As for the usage of content features, \textbf{PP-Rec} uses a similar news encoder as the one from \textbf{NRMS}, except with additional entity information and cross-attention operations (the details of the news encoder are abstracted in the figure for the sake of space). \textbf{PP-Rec} incorporates contextual features mainly in two ways. First, it proposes a news popularity predictor (left side of the figure), which takes CTR, freshness, and content features as its input. The CTR is used as its raw value, while freshness is quantized into an embedding, and content features are processed by a separate news encoder from the one in the content side. The last two are combined with a gating operation, whose output is combined with CTR by weighted summation to produce the final popularity prediction. The second usage of contextual features in \textbf{PP-Rec} is in its user encoder (right side of the figure), which is also similar to the one from \textbf{NRMS}, except for the additional usage of CTR value and content-popularity joint attention network (the details of the user encoder are also abstracted in the figure). In this module, CTR value is quantized into an embedding to be used in the joint attention network. Finally, \textbf{PP-Rec} aggregates the results from the popularity predictor and from the content side via a personalized aggregator that uses the computed user embedding as the input for the gating operation.

Since \textbf{PP-Rec} requires CTR and freshness features, experiments couldn't be conducted on \textbf{MIND}, \textbf{Globo} and \textbf{Adressa} \footnote{The authors of \textbf{PP-Rec} recognized this problem and had to build their own proprietary datasets containing these features. However, to our knowledge, those datasets weren't published.}. In addition, since it also requires entity information, it also couldn't be experimented on \textbf{Prop}. However, this would mean eliminating \textbf{PP-Rec} from our experiments completely, which could be viewed as an unfair comparison. Therefore, we enabled its experiments for \textbf{Prop} by using its variation where we removed entity inputs and the corresponding cross-attentions in its news encoder, which basically makes it identical to the one from \textbf{NRMS}.

Third, naive context models are the ones that utilize the features we've defined in Section \ref{feature_definitions} without applying any sort of operations to them. For example, \textbf{Naive-CTR} will rank the candidate news articles just by their $CTR(t)$ values (articles with higher $CTR(t)$ values will be ranked higher). Its concept is captured in Figure \ref{fig:five_models} (c).
Likewise, \textbf{Naive-numclicks}, \textbf{Naive-trendiness}, and \textbf{Naive-freshness} will rank the items according to their $numclicks(t)$, $trendiness(t)$, and $freshness(t)$ \footnote{In the case of $freshness$, lower values will be ranked higher because it implies that the article is more fresh}. These models will serve as baselines that show the bare effects of contextual features alone.

Fourth, for deep-learning content models with naive context combination, we took \textbf{NRMS} and simply added the contextual features to its side. Its structure is illustrated in Figure \ref{fig:five_models} (d).
In \textbf{NRMS+Naive}, the logits from \textbf{NRMS} are combined with contextual features of candidate items by a weighted sum with learnable weights. Since $numclicks$, $trendiness$, and $freshness$ offer values that are intolerably high for neural networks, we applied $log(1+x)$ operation to scale each of them following \cite{newspersonalizationusingsupportvectormachines}. However, except for these scaling operations, note that we omitted any other non-linear operations such as MLPs (Multi-Layer Perceptrons) \footnote{Hence the term \textbf{NRMS+Naive}}. Also note that while \textbf{PP-Rec} uses the contextual features in an indirect way (through sophisticated operations such as embedding, gating, cross-attentions, etc), in \textbf{NRMS-Naive} we simply use them as logits that directly contribute to the recommendation prediction.

Finally, for gradient boosting context models, we organized the contextual features and labels (i.e. clicks) into a tabular form. Then, following \cite{hyperconnectboosting}, we used CatBoost library \cite{catboost} to build a prediction model from these features. This model serves to show how much more can be exploited from the contextual features when we apply learnable non-linear operations such as decision trees.

%% file: sections/boosting_subsections/results_table.tex
\begin{table}[t]
\caption{Performance Comparison between four kinds of models. The numbers are measured by NDCG@10 metric.}
\Description[Description]{Description}
\label{tab:context_performance_comparison}

\begin{tabular}{@{}l|cccc@{}}
\toprule
 &  \textbf{Prop} &  \textbf{Mind} &  \textbf{Globo} &  \textbf{Adressa} \\
 \midrule
\textbf{NRMS} &  0.5327 & 0.3712 &  0.3595 &  0.3467 \\
 \midrule
 \textbf{PP-Rec} & 0.5286 & - & - & - \\
 \midrule
\textbf{Naive-CTR}  &  0.5858 &  0.4023 &  - &  - \\
\textbf{Naive-numclicks} &  0.5012 &  0.3439 &  0.2467 &  0.2041 \\
\textbf{Naive-trendiness} &  0.5195 &  0.3744 &  0.2642 &  0.2947 \\
\textbf{Naive-freshness} &  0.35 &  - &  0.2822 &  0.3107 \\
 \midrule
\textbf{NRMS+Naive} &  0.5985 &  0.3946 &  0.3697 &  0.3832 \\
 \midrule
\textbf{Gradient Boosting} &  0.5865 &  0.4034 &  0.5198 &  0.4281 \\
\bottomrule
\end{tabular}



\end{table}

%% file: sections/boosting_subsections/pprec_ablation_table.tex
\begin{table}[t]
\caption{\textbf{PP-Rec} ablation.}
\label{tab:pprec_ablation}
\begin{tabular}{@{}l|c|c|c|c@{}}
    \toprule
        \textbf{Model} & \textbf{Recall@10} & \textbf{NDCG@10} \\
    \midrule
        \textbf{PP-Rec} & 0.7963 & 0.5286 \\
        \textbf{PP-Rec-onlyctr} & 0.798 & 0.5312 \\
        \textbf{PP-Rec-onlyctr-noagg} & 0.8315 & 0.5944 \\
        \textbf{NRMS+Naive} & 0.8291 & 0.5985 \\
    \bottomrule
\end{tabular}
\end{table}

%% file: sections/boosting_subsections/results.tex
Table \ref{tab:context_performance_comparison} shows the performance comparison between the five different kinds of methods introduced earlier. Many things are noticeable in this comparison.

First is the surprising effectiveness of CTR. Note that \textbf{Naive-CTR} outperformed \textbf{NRMS} in both datasets where $CTR(t)$ feature was available. This is very impressive because \textbf{Naive-CTR} required no training at all, and $CTR(t)$ feature wasn't transformed by any operations. This suggests that $CTR(t)$, a global measure of the amount of appeal that the article has to the population at time $t$, can be a strong indicator of whether the next user will click it or not. It even suggests that this global context feature could be a stronger cause of a click than the personal preferences of the user, which is understandable when considering the nature of news media.

Second, for similar reasons as the first, other context features also proved to be effective, offering decent performance in \textbf{Prop} and \textbf{Mind} without any kind of operations. $numclicks(t)$, which is equivalent to the numerator in the definition of $CTR(t)$, proved to be less effective than $CTR(t)$, which again emphasizes the importance of having impression data in the dataset. $trendiness(t)$, which incorporate temporal factors into $numclicks(t)$, proved to be more effective, which suggests that temporal aspects must indeed be considered. $freshness(t)$ was shown to be the most effective in \textbf{Globo} and \textbf{Adressa}, while it wasn't shown to be a defining factor in \textbf{Prop}.

Third, the performance gap between \textbf{PP-Rec} and \textbf{NRMS+Naive} is noticeable. Recall that while they utilize the same features (both content and context), \textbf{PP-Rec} utilizes the contextual feautres in an intricate manner while \textbf{NRMS+Naive} just uses them as additional logits. This seems to contradict the myth that complicated deep-learning operations must be applied to the features in order to utilize them properly. In fact, to investigate the factors that were hindering the performance of \textbf{PP-Rec}, we conducted a simple ablation study by taking out some of its components.
Table \ref{tab:pprec_ablation} shows the performance comparison between \textbf{PP-Rec} and two of its variants.
The first variant, \textbf{PP-Rec-onlyctr}, was created by removing freshness and content components in the news popularity predictor module, leaving only the CTR value as its output. As a result, the performance increased slightly, but the big gap persisted. Then, we identified the big difference between \textbf{PP-Rec-onlyctr} and \textbf{NRMS+Naive}, which was the personalized aggregator module at the final layer in \textbf{PP-Rec}, which was blocking the direct contribution of CTR value to the final prediction. Replacing this layer with a simple linear one as in \textbf{NRMS+Naive} (\textbf{PP-Rec-onlyctr-noagg}), the performance greatly increased and achieved similar performance as \textbf{NRMS+Naive}. This suggests that direct usage of contextual features can be more effective than using them indirectly through complex deep-learning operations.

Fourth, the impressive performance of gradient boosting model is notable. In \textbf{Mind}, \textbf{Globo}, and \textbf{Adressa}, gradient boosting method outperforms \textbf{NRMS+Naive}. This is very impressive because the former only utilizes the contextual features with a light tabular method, whereas the latter utilizes content features besides the contextual features, with a deep-learning method that is much heavier than gradient boosting both in terms of the number of parameters and the training time. Since \textbf{NRMS+Naive} uses contextual features without any operations, whereas the gradient boosting model uses them via decision trees, we can infer that the presence of such learnable non-linear operations are crucial for good recommendation performance.

Lastly, we stress the importance of good quality datasets to facilitate news recommendation research. Note that the absence of some features in the datasets crucially affected the experiments on multiple levels. On the feature-level, the absence of impression data made it impossible to compute features such as CTR, which excludes the possibility of utilizing these features for recommendation. Even for the datasets where impression data was present, note that the $CTR(t)$ values we used were just an approximation, since it was reconstructed from only within the dataset which is only a sample from the real-world. The same goes for every other contextual features. Therefore, better results are possible for those that have access to the real-world records, or at least the true values for the contextual features.
On the evaluation-level, for datasets with no impression items, negative items couldn't be constructed naturally, and instead had to be sampled from other users' clicks. This makes the experiments inaccurate, since it is difficult to regard those items as being truly negative because the users might have clicked the items had they been shown to them. In fact, this factor could have contributed to the comparatively large performance gap in \textbf{Globo} and \textbf{Adressa} compared to \textbf{Prop} and \textbf{Mind}. Also, since \textbf{Globo} provides no text information, existing content deep-learning models such as \textbf{NRMS} couldn't be used as-is, and had to be transformed into a weaker version. We assume that the unusually large performance gap in \textbf{Globo} results from this tragedy. These observations offer a consideration for future parties willing to publish a good quality news recommendation dataset.


We summarize the lessons of this section below.

\begin{itemize}
  \item Contextual features such as CTR, popularity, and freshness prove to be very powerful. Without any operations, contextual features alone can match and sometimes outperform the heavy deep-learning models that use content features.
  \item Contextual features have largely been neglected or underutilized in recent models.
  \item Recent models that tried to incorporate contextual features haven't been able to use them to their full potential. Our analysis shows that such underutilization results from trying to use them indirectly through sophisticated deep-learning operations, whereas a more direct approach proved to be more effective.
  \item When using these contextual features directly, having a learnable non-linear operation applied to them is crucial for recommendation performance. This is shown by the effectiveness of the gradient boosting method.
  \item Many published news recommendation datasets lack crucial features such as impression data. This affects fair evaluation of certain news recommendation models and can possibly hinder the research in this area.
\end{itemize}

%% file: sections/deeplearning.tex
\subsection{Motivation}
\input{sections/deeplearning_subsections/motivation}

\subsection{Method}
\input{sections/deeplearning_subsections/method}



\subsection{Experiments}
\input{sections/deeplearning_subsections/experiments}

\subsection{Results}
\input{sections/deeplearning_subsections/results}

\input{sections/deeplearning_subsections/results_table}

%% file: sections/deeplearning_subsections/motivation.tex
Although the previous section seems to imply that the contextual features and cheap tabular methods are all we need for good news recommendation, this is not quite the case when the data size increases. Recall that the previous experiments were each done with a subset of 5k users. As the data size grows into a larger scale, it becomes difficult to outperform deep-learning content models by contextual features and cheap tabular methods alone (however, the comparison and the analysis from the previous section still hold at this scale). In addition, while it is true that non-personal features such as CTR or popularity greatly affect the users' news consumption, the users' distinct preferences cannot be ignored in order to provide personalized news recommendation, thereby making the usage of content features necessary. Furthermore, since many companies already deploy deep-learning based models, it would be more convenient for them to reinforce their models if our solution is in the form of a deep-learning module.

For these reasons, we suggest a contextual deep-learning module, which can be attached to any previous content-based deep-learning model in a modular manner to boost their performance by a large margin.

%% file: sections/deeplearning_subsections/method.tex
We propose a purposefully simple method to accommodate the lessons learned in the previous section. Since direct usage of contextual features proved to be more effective, we suggest to use the architecture illustrated in Figure \ref{fig:five_models} (d). The left side of the figure represents any previous content-based deep-learning news recommendation model, while the right side of the figure represents an additional context module that we propose. Unlike Figure \ref{fig:five_models} (d) however, we suggest to apply an additional dense layer (i.e. MLP) right after the scaling operation of the contextual features. These dense layers, with learnable parameters, offer the non-linearity that was shown to be crucial in the previous section. Note that other learnable non-linear operations such as decision trees are still applicable here, but it would require an additional engineering to connect the deep-learning components and the decision tree. Therefore, for companies that are already deploying deep-learning recommendation models at scale, our approach comes in handy.







%% file: sections/deeplearning_subsections/experiments.tex
To show the effectiveness of our method, we conducted the following experiment. Using the same datasets and evaluation schemes in Section \ref{section4}, we evaluated how much the recommendation performance is increased when we attach our context module to a previous content-based model. For this experiment we used the full-scale datasets to give the deep-learning models the full leverage(?). We used the following models as baselines:
\input{sections/deeplearning_subsections/baselines}

%% file: sections/deeplearning_subsections/baselines.tex
\begin{itemize}
\item \textbf{NPA}~\cite{NPA}: A news recommendation model that uses CNNs and personalized attention module to identify important words with regard to each user in the news encoder. Words from news titles are used as news encoder inputs. Similarly, personalized attention is used as user encoder to identify important articles with regard to each user.
\item \textbf{NRMS}~\cite{NRMS}: A news recommendation model that uses self-attention and additive attention operation to learn both news and user representations. Words from news titles are used as news encoder inputs.
\item \textbf{LSTUR}~\cite{an2019}: A news recommendation model that uses GRUs and ID-based embeddings to learn the user's short-term and long-term representations respectively. It uses CNNs and additive attentions on word embeddings then concatenate the output with (sub)category embeddings to represent an article.
\item \textbf{NAML}~\cite{NAML}: A news recommendation model that uses CNNs and additive attentions on word embeddings and applies another additive attention on its output together with (sub)category embeddings to learn news representations. It uses additive attention operation to learn user representations.
\end{itemize}

We implemented all baselines in PyTorch \cite{pytorch} and experimented with the same device (Tesla M40). For each model we used the hyperparameters suggested by the authors. We used appropriate pre-trained word embeddings for each dataset (300-dimension Glove embeddings for \textbf{Mind} \footnote{https://nlp.stanford.edu/projects/glove/}, 100-dimension skip-gram Norwegian word embeddings for \textbf{Adressa} \footnote{Model \#100 in http://vectors.nlpl.eu/repository/.}, and 128-dimension word embeddings pre-trained on large corpus of proprietary news articles for \textbf{Prop}).

As in our earlier experiment, for \textbf{Globo}, since it doesn't provide articles' textual information and instead provide their pre-trained item embeddings, we had to use variations of the models where we used the pre-trained item embeddings in place of news encoder outputs.

%% file: sections/deeplearning_subsections/results.tex
Table \ref{tab:performance_comparison} shows the comparison results. As the table shows, the addition of our context module increases the performance of previous content-based models by a large margin. As noted earlier, the relatively larger gaps in \textbf{Globo} and \textbf{Adressa} could have partially resulted from their lack of crucial features. However, since the experiments in \textbf{Prop} and \textbf{Mind} were done in a more realistic setting, our module can be said to be effective in a real-world environment.

Note that the performance gain is profitably high, especially when considering that the addition of our context module adds very small number of parameters and the amount of computation to the system compared to the heavy content models.

%% file: sections/deeplearning_subsections/results_table.tex

\begin{table*}[]
\caption{Performance Comparison. +CTX columns denote the model on its left side with our context module attached to it. Each "Improv." column computes the relative improvement between the two.}
\Description[Description]{Description}
\label{tab:performance_comparison}

\begin{tabular}{@{}ll|ccc|ccc|ccc|ccc@{}}
\toprule
Dataset & Metric &  NPA &  +CTX &  Improv. &  NRMS &  +CTX &  Improv. &  LSTUR &  +CTX &  Improv. &  NAML &  +CTX &  Improv. \\

\midrule
\multirow{3}{*}{Prop} & AUC & 0.6716 & 0.7097 & \textbf{5.67\%} & 0.6773 & 0.7299 & \textbf{7.76\%} & 0.6680 & 0.7209 & \textbf{7.91\%} & 0.6865 & 0.7287 & \textbf{6.14\%} \\
  & Recall@10 & 0.7999 & 0.8218 & \textbf{2.74\%} & 0.8049 & 0.8337 & \textbf{3.59\%} & 0.7963 & 0.8251 & \textbf{3.63\%} & 0.8078 & 0.8341 & \textbf{3.27\%} \\
  & NDCG@10 & 0.5366 & 0.5766 & \textbf{7.45\%} & 0.5418 & 0.5954 & \textbf{9.91\%} & 0.5328 & 0.5841 & \textbf{9.63\%} & 0.5512 & 0.5936 & \textbf{7.70\%} \\
\midrule
\multirow{3}{*}{Mind} & AUC & 0.6509 & 0.6943 & \textbf{6.66\%} & 0.6636 & 0.7086 & \textbf{6.78\%} & 0.6793 & 0.7094 & \textbf{4.42\%} & 0.6787 & 0.7083 & \textbf{4.36\%} \\
  & Recall@10 & 0.6474 & 0.6799 & \textbf{5.01\%} & 0.6595 & 0.6948 & \textbf{5.35\%} & 0.6691 & 0.6939 & \textbf{3.72\%} & 0.6776 & 0.6971 & \textbf{2.89\%} \\
  & NDCG@10 & 0.3857 & 0.4304 & \textbf{11.60\%} & 0.4020 & 0.4439 & \textbf{10.43\%} & 0.4154 & 0.4476 & \textbf{7.75\%} & 0.4191 & 0.4496 & \textbf{7.28\%} \\
\midrule
\multirow{3}{*}{Globo} & AUC & 0.6390 & 0.6306 & \textbf{-1.32\%} & 0.6431 & 0.7106 & \textbf{10.49\%} & 0.6503 & 0.7118 & \textbf{9.46\%} & 0.6558 & 0.7156 & \textbf{9.12\%} \\
  & Recall@10 & 0.6503 & 0.6150 & \textbf{-5.43\%} & 0.6562 & 0.7281 & \textbf{10.96\%} & 0.6620 & 0.7306 & \textbf{10.36\%} & 0.6723 & 0.7324 & \textbf{8.94\%} \\
  & NDCG@10 & 0.3746 & 0.3849 & \textbf{2.75\%} & 0.3768 & 0.4673 & \textbf{24.02\%} & 0.3919 & 0.4701 & \textbf{19.95\%} & 0.3975 & 0.4763 & \textbf{19.82\%} \\
\midrule
\multirow{3}{*}{Adressa} & AUC & 0.6265 & 0.6756 & \textbf{7.84\%} & 0.6306 & 0.6800 & \textbf{7.83\%} & 0.6888 & 0.7182 & \textbf{4.27\%} & 0.6441 & 0.6975 & \textbf{8.29\%} \\
  & Recall@10 & 0.6290 & 0.6726 & \textbf{6.93\%} & 0.6422 & 0.6809 & \textbf{6.03\%} & 0.7093 & 0.7319 & \textbf{3.19\%} & 0.6561 & 0.7045 & \textbf{7.38\%} \\
  & NDCG@10 & 0.3763 & 0.4397 & \textbf{16.85\%} & 0.3737 & 0.4463 & \textbf{19.43\%} & 0.4367 & 0.4868 & \textbf{11.47\%} & 0.3855 & 0.4669 & \textbf{21.12\%} \\
\bottomrule
\end{tabular}


\end{table*}

%% file: sections/conclusion.tex
In this paper we sought to reevaluate the significance of contextual features of news articles. We conducted a thorough comparison between recent deep-learning news recommendation models and naive contextual models on four real-world datasets. As a result, we learned (1) that contextual features hold great potential for improving recommendation performance, (2) that these features were largely neglected or underutilized in recent research, (3) that one reason for such underutilization was the tendency to apply overly sophisticated deep-learning operations to the features, (4) that a more direct usage with learnable non-linear operations prove to be much more effective, and (5) that currently published news recommendation datasets lack crucial features that hinder fair evaluation of contextual models. From this knowledge, we proposed a simple contextual deep-learning module that can be attached to previous content-based solutions in a modular manner.
Experiments showed that our module boosts the recommendation performance by a large margin on many real-world datasets.

Although in this paper we only proposed a context module for news articles, suggesting a similar module for users could be a possible future work. The user's context might include: user's CTR, user group, age, gender, location, etc. As \cite{modelagnosticcounterfactual} pointed out, the separation of user module, item module, and user-item module could enable a counterfactual inference and reduce the popularity bias. Also, as mentioned earlier, examining more contextual features such as user-group specific CTR remains as a future work.

%% file: main.bbl

\begin{thebibliography}{69}


\ifx \showCODEN    \undefined \def \showCODEN     #1{\unskip}     \fi
\ifx \showDOI      \undefined \def \showDOI       #1{#1}\fi
\ifx \showISBNx    \undefined \def \showISBNx     #1{\unskip}     \fi
\ifx \showISBNxiii \undefined \def \showISBNxiii  #1{\unskip}     \fi
\ifx \showISSN     \undefined \def \showISSN      #1{\unskip}     \fi
\ifx \showLCCN     \undefined \def \showLCCN      #1{\unskip}     \fi
\ifx \shownote     \undefined \def \shownote      #1{#1}          \fi
\ifx \showarticletitle \undefined \def \showarticletitle #1{#1}   \fi
\ifx \showURL      \undefined \def \showURL       {\relax}        \fi
\providecommand\bibfield[2]{#2}
\providecommand\bibinfo[2]{#2}
\providecommand\natexlab[1]{#1}
\providecommand\showeprint[2][]{arXiv:#2}

\bibitem[\protect\citeauthoryear{An, Wu, Wu, Zhang, Liu, and Xie}{An
  et~al\mbox{.}}{2019}]%
        {an2019}
\bibfield{author}{\bibinfo{person}{Mingxiao An}, \bibinfo{person}{Fangzhao Wu},
  \bibinfo{person}{Chuhan Wu}, \bibinfo{person}{Kun Zhang},
  \bibinfo{person}{Zheng Liu}, {and} \bibinfo{person}{Xing Xie}.}
  \bibinfo{year}{2019}\natexlab{}.
\newblock \showarticletitle{Neural news recommendation with long-and short-term
  user representations}. In \bibinfo{booktitle}{\emph{Proceedings of the 57th
  Annual Meeting of the Association for Computational Linguistics}}.
  \bibinfo{pages}{336--345}.
\newblock


\bibitem[\protect\citeauthoryear{Chen and Guestrin}{Chen and Guestrin}{2016}]%
        {xgboost}
\bibfield{author}{\bibinfo{person}{Tianqi Chen} {and} \bibinfo{person}{Carlos
  Guestrin}.} \bibinfo{year}{2016}\natexlab{}.
\newblock \showarticletitle{{XGBoost}: A Scalable Tree Boosting System}. In
  \bibinfo{booktitle}{\emph{Proceedings of the 22nd ACM SIGKDD International
  Conference on Knowledge Discovery and Data Mining}} (San Francisco,
  California, USA) \emph{(\bibinfo{series}{KDD '16})}.
  \bibinfo{publisher}{ACM}, \bibinfo{address}{New York, NY, USA},
  \bibinfo{pages}{785--794}.
\newblock
\showISBNx{978-1-4503-4232-2}
\urldef\tempurl%
\url{https://doi.org/10.1145/2939672.2939785}
\showDOI{\tempurl}


\bibitem[\protect\citeauthoryear{Chu and Park}{Chu and Park}{2009}]%
        {dynamicbilinear}
\bibfield{author}{\bibinfo{person}{Wei Chu} {and} \bibinfo{person}{Seung-Taek
  Park}.} \bibinfo{year}{2009}\natexlab{}.
\newblock \showarticletitle{Personalized recommendation on dynamic content
  using predictive bilinear models}. In \bibinfo{booktitle}{\emph{Proceedings
  of the 18th international conference on World wide web}}.
  \bibinfo{pages}{691--700}.
\newblock


\bibitem[\protect\citeauthoryear{Dai, Yang, Yang, Carbonell, Le, and
  Salakhutdinov}{Dai et~al\mbox{.}}{2019}]%
        {TRANSFORMER-XL}
\bibfield{author}{\bibinfo{person}{Zihang Dai}, \bibinfo{person}{Zhilin Yang},
  \bibinfo{person}{Yiming Yang}, \bibinfo{person}{Jaime~G. Carbonell},
  \bibinfo{person}{Quoc~Viet Le}, {and} \bibinfo{person}{Ruslan
  Salakhutdinov}.} \bibinfo{year}{2019}\natexlab{}.
\newblock \showarticletitle{Transformer-XL: Attentive Language Models beyond a
  Fixed-Length Context}.
\newblock \bibinfo{journal}{\emph{Association for Computational Linguistics
  (ACL)}} (\bibinfo{year}{2019}).
\newblock


\bibitem[\protect\citeauthoryear{Darvishy, Ibrahim, Sidi, and
  Mustapha}{Darvishy et~al\mbox{.}}{2020}]%
        {hypner}
\bibfield{author}{\bibinfo{person}{Asghar Darvishy}, \bibinfo{person}{Hamidah
  Ibrahim}, \bibinfo{person}{Fatimah Sidi}, {and} \bibinfo{person}{Aida
  Mustapha}.} \bibinfo{year}{2020}\natexlab{}.
\newblock \showarticletitle{HYPNER: A Hybrid Approach for Personalized News
  Recommendation}.
\newblock \bibinfo{journal}{\emph{IEEE Access}}  \bibinfo{volume}{8}
  (\bibinfo{year}{2020}), \bibinfo{pages}{46877--46894}.
\newblock


\bibitem[\protect\citeauthoryear{Das, Datar, Garg, and Rajaram}{Das
  et~al\mbox{.}}{2007}]%
        {das2007google}
\bibfield{author}{\bibinfo{person}{Abhinandan~S Das}, \bibinfo{person}{Mayur
  Datar}, \bibinfo{person}{Ashutosh Garg}, {and} \bibinfo{person}{Shyam
  Rajaram}.} \bibinfo{year}{2007}\natexlab{}.
\newblock \showarticletitle{Google news personalization: scalable online
  collaborative filtering}. In \bibinfo{booktitle}{\emph{Proceedings of the
  16th international conference on World Wide Web}}. \bibinfo{pages}{271--280}.
\newblock


\bibitem[\protect\citeauthoryear{de~Souza Pereira~Moreira, Ferreira, and
  da~Cunha}{de~Souza Pereira~Moreira et~al\mbox{.}}{2018}]%
        {globodataset}
\bibfield{author}{\bibinfo{person}{Gabriel de Souza Pereira~Moreira},
  \bibinfo{person}{Felipe Ferreira}, {and} \bibinfo{person}{Adilson~Marques da
  Cunha}.} \bibinfo{year}{2018}\natexlab{}.
\newblock \showarticletitle{News session-based recommendations using deep
  neural networks}. In \bibinfo{booktitle}{\emph{Proceedings of the 3rd
  Workshop on Deep Learning for Recommender Systems}}. \bibinfo{pages}{15--23}.
\newblock


\bibitem[\protect\citeauthoryear{Devlin, Chang, Lee, and Toutanova}{Devlin
  et~al\mbox{.}}{2019}]%
        {BERT}
\bibfield{author}{\bibinfo{person}{Jacob Devlin}, \bibinfo{person}{Ming{-}Wei
  Chang}, \bibinfo{person}{Kenton Lee}, {and} \bibinfo{person}{Kristina
  Toutanova}.} \bibinfo{year}{2019}\natexlab{}.
\newblock \showarticletitle{{BERT:} Pre-training of Deep Bidirectional
  Transformers for Language Understanding}.
\newblock \bibinfo{journal}{\emph{the North American Chapter of the Association
  for Computational Linguistics: Human Language Technologies (NAACL-HLT)}}
  (\bibinfo{year}{2019}).
\newblock


\bibitem[\protect\citeauthoryear{Dorogush, Ershov, and Gulin}{Dorogush
  et~al\mbox{.}}{2018}]%
        {catboost}
\bibfield{author}{\bibinfo{person}{Anna~Veronika Dorogush},
  \bibinfo{person}{Vasily Ershov}, {and} \bibinfo{person}{Andrey Gulin}.}
  \bibinfo{year}{2018}\natexlab{}.
\newblock \showarticletitle{CatBoost: gradient boosting with categorical
  features support}.
\newblock \bibinfo{journal}{\emph{arXiv preprint arXiv:1810.11363}}
  (\bibinfo{year}{2018}).
\newblock


\bibitem[\protect\citeauthoryear{Garcin and Faltings}{Garcin and
  Faltings}{2013}]%
        {penrecsys}
\bibfield{author}{\bibinfo{person}{Florent Garcin} {and} \bibinfo{person}{Boi
  Faltings}.} \bibinfo{year}{2013}\natexlab{}.
\newblock \showarticletitle{Pen recsys: A personalized news recommender systems
  framework}. In \bibinfo{booktitle}{\emph{Proceedings of the 2013
  International News Recommender Systems Workshop and Challenge}}.
  \bibinfo{pages}{3--9}.
\newblock


\bibitem[\protect\citeauthoryear{Ge, Wu, Wu, Qi, and Huang}{Ge
  et~al\mbox{.}}{2020}]%
        {ge2020graph}
\bibfield{author}{\bibinfo{person}{Suyu Ge}, \bibinfo{person}{Chuhan Wu},
  \bibinfo{person}{Fangzhao Wu}, \bibinfo{person}{Tao Qi}, {and}
  \bibinfo{person}{Yongfeng Huang}.} \bibinfo{year}{2020}\natexlab{}.
\newblock \showarticletitle{Graph enhanced representation learning for news
  recommendation}. In \bibinfo{booktitle}{\emph{Proceedings of The Web
  Conference 2020}}. \bibinfo{pages}{2863--2869}.
\newblock


\bibitem[\protect\citeauthoryear{Gershman, Wolfe, Fink, and Carbonell}{Gershman
  et~al\mbox{.}}{2011}]%
        {newspersonalizationusingsupportvectormachines}
\bibfield{author}{\bibinfo{person}{Anatole Gershman}, \bibinfo{person}{Travis
  Wolfe}, \bibinfo{person}{Eugene Fink}, {and} \bibinfo{person}{Jaime~G
  Carbonell}.} \bibinfo{year}{2011}\natexlab{}.
\newblock \showarticletitle{News personalization using support vector
  machines}.
\newblock  (\bibinfo{year}{2011}).
\newblock


\bibitem[\protect\citeauthoryear{Gulla, Zhang, Liu, {\"O}zg{\"o}bek, and
  Su}{Gulla et~al\mbox{.}}{2017}]%
        {adressadataset}
\bibfield{author}{\bibinfo{person}{Jon~Atle Gulla}, \bibinfo{person}{Lemei
  Zhang}, \bibinfo{person}{Peng Liu}, \bibinfo{person}{{\"O}zlem
  {\"O}zg{\"o}bek}, {and} \bibinfo{person}{Xiaomeng Su}.}
  \bibinfo{year}{2017}\natexlab{}.
\newblock \showarticletitle{The adressa dataset for news recommendation}. In
  \bibinfo{booktitle}{\emph{Proceedings of the international conference on web
  intelligence}}. \bibinfo{pages}{1042--1048}.
\newblock


\bibitem[\protect\citeauthoryear{Hidasi and Karatzoglou}{Hidasi and
  Karatzoglou}{2018}]%
        {hidasi2018recurrent}
\bibfield{author}{\bibinfo{person}{Bal{\'a}zs Hidasi} {and}
  \bibinfo{person}{Alexandros Karatzoglou}.} \bibinfo{year}{2018}\natexlab{}.
\newblock \showarticletitle{Recurrent neural networks with top-k gains for
  session-based recommendations}.
\newblock \bibinfo{journal}{\emph{International Conference on Information and
  Knowledge Management (CIKM)}} (\bibinfo{year}{2018}).
\newblock


\bibitem[\protect\citeauthoryear{Hidasi, Karatzoglou, Baltrunas, and
  Tikk}{Hidasi et~al\mbox{.}}{2016}]%
        {GRU4Rec}
\bibfield{author}{\bibinfo{person}{Bal{\'{a}}zs Hidasi},
  \bibinfo{person}{Alexandros Karatzoglou}, \bibinfo{person}{Linas Baltrunas},
  {and} \bibinfo{person}{Domonkos Tikk}.} \bibinfo{year}{2016}\natexlab{}.
\newblock \showarticletitle{Session-based Recommendations with Recurrent Neural
  Networks}.
\newblock \bibinfo{journal}{\emph{International Conference on Learning
  Representations (ICLR)}} (\bibinfo{year}{2016}).
\newblock


\bibitem[\protect\citeauthoryear{Huang, He, Gao, Deng, Acero, and Heck}{Huang
  et~al\mbox{.}}{2013}]%
        {DSSM}
\bibfield{author}{\bibinfo{person}{Po-Sen Huang}, \bibinfo{person}{Xiaodong
  He}, \bibinfo{person}{Jianfeng Gao}, \bibinfo{person}{Li Deng},
  \bibinfo{person}{Alex Acero}, {and} \bibinfo{person}{Larry Heck}.}
  \bibinfo{year}{2013}\natexlab{}.
\newblock \showarticletitle{Learning deep structured semantic models for web
  search using clickthrough data}. In \bibinfo{booktitle}{\emph{Proceedings of
  the 22nd ACM international conference on Information \& Knowledge
  Management}}. \bibinfo{pages}{2333--2338}.
\newblock


\bibitem[\protect\citeauthoryear{Ilievski and Roy}{Ilievski and Roy}{2013}]%
        {personalizednewsrecommendationbasedonimplicitfeedback}
\bibfield{author}{\bibinfo{person}{Ilija Ilievski} {and} \bibinfo{person}{Sujoy
  Roy}.} \bibinfo{year}{2013}\natexlab{}.
\newblock \showarticletitle{Personalized news recommendation based on implicit
  feedback}. In \bibinfo{booktitle}{\emph{Proceedings of the 2013 international
  news recommender systems workshop and challenge}}. \bibinfo{pages}{10--15}.
\newblock


\bibitem[\protect\citeauthoryear{Jankiewicz, Kyrashchuk, Sienkowski, and
  W{\'o}jcik}{Jankiewicz et~al\mbox{.}}{2019}]%
        {recsys2019challengewinner}
\bibfield{author}{\bibinfo{person}{Pawe{\l} Jankiewicz},
  \bibinfo{person}{Liudmyla Kyrashchuk}, \bibinfo{person}{Pawe{\l} Sienkowski},
  {and} \bibinfo{person}{Magdalena W{\'o}jcik}.}
  \bibinfo{year}{2019}\natexlab{}.
\newblock \showarticletitle{Boosting algorithms for a session-based,
  context-aware recommender system in an online travel domain}. In
  \bibinfo{booktitle}{\emph{Proceedings of the Workshop on ACM Recommender
  Systems Challenge}}. \bibinfo{pages}{1--5}.
\newblock


\bibitem[\protect\citeauthoryear{Jonnalagedda and Gauch}{Jonnalagedda and
  Gauch}{2013}]%
        {twitterpopularity}
\bibfield{author}{\bibinfo{person}{Nirmal Jonnalagedda} {and}
  \bibinfo{person}{Susan Gauch}.} \bibinfo{year}{2013}\natexlab{}.
\newblock \showarticletitle{Personalized news recommendation using twitter}. In
  \bibinfo{booktitle}{\emph{2013 IEEE/WIC/ACM International Joint Conferences
  on Web Intelligence (WI) and Intelligent Agent Technologies (IAT)}},
  Vol.~\bibinfo{volume}{3}. IEEE, \bibinfo{pages}{21--25}.
\newblock


\bibitem[\protect\citeauthoryear{Jonnalagedda, Gauch, Labille, and
  Alfarhood}{Jonnalagedda et~al\mbox{.}}{2016}]%
        {incorporatingpopularity}
\bibfield{author}{\bibinfo{person}{Nirmal Jonnalagedda}, \bibinfo{person}{Susan
  Gauch}, \bibinfo{person}{Kevin Labille}, {and} \bibinfo{person}{Sultan
  Alfarhood}.} \bibinfo{year}{2016}\natexlab{}.
\newblock \showarticletitle{Incorporating popularity in a personalized news
  recommender system}.
\newblock \bibinfo{journal}{\emph{PeerJ Computer Science}}  \bibinfo{volume}{2}
  (\bibinfo{year}{2016}), \bibinfo{pages}{e63}.
\newblock


\bibitem[\protect\citeauthoryear{{Kang} and {McAuley}}{{Kang} and
  {McAuley}}{2018}]%
        {SASRec}
\bibfield{author}{\bibinfo{person}{W. {Kang}} {and} \bibinfo{person}{J.
  {McAuley}}.} \bibinfo{year}{2018}\natexlab{}.
\newblock \showarticletitle{Self-Attentive Sequential Recommendation}.
\newblock \bibinfo{journal}{\emph{International Conference on Data Mining
  (ICDM)}} (\bibinfo{year}{2018}).
\newblock


\bibitem[\protect\citeauthoryear{Kazai, Yusof, and Clarke}{Kazai
  et~al\mbox{.}}{2016}]%
        {facebooktwitterlocation}
\bibfield{author}{\bibinfo{person}{Gabriella Kazai}, \bibinfo{person}{Iskander
  Yusof}, {and} \bibinfo{person}{Daoud Clarke}.}
  \bibinfo{year}{2016}\natexlab{}.
\newblock \showarticletitle{Personalised news and blog recommendations based on
  user location, Facebook and Twitter user profiling}. In
  \bibinfo{booktitle}{\emph{Proceedings of the 39th International ACM SIGIR
  conference on Research and Development in Information Retrieval}}.
  \bibinfo{pages}{1129--1132}.
\newblock


\bibitem[\protect\citeauthoryear{Ke, Meng, Finley, Wang, Chen, Ma, Ye, and
  Liu}{Ke et~al\mbox{.}}{2017}]%
        {lightgbm}
\bibfield{author}{\bibinfo{person}{Guolin Ke}, \bibinfo{person}{Qi Meng},
  \bibinfo{person}{Thomas Finley}, \bibinfo{person}{Taifeng Wang},
  \bibinfo{person}{Wei Chen}, \bibinfo{person}{Weidong Ma},
  \bibinfo{person}{Qiwei Ye}, {and} \bibinfo{person}{Tie-Yan Liu}.}
  \bibinfo{year}{2017}\natexlab{}.
\newblock \showarticletitle{Lightgbm: A highly efficient gradient boosting
  decision tree}.
\newblock \bibinfo{journal}{\emph{Advances in neural information processing
  systems}}  \bibinfo{volume}{30} (\bibinfo{year}{2017}),
  \bibinfo{pages}{3146--3154}.
\newblock


\bibitem[\protect\citeauthoryear{Khattar, Kumar, Varma, and Gupta}{Khattar
  et~al\mbox{.}}{2018a}]%
        {HRAM}
\bibfield{author}{\bibinfo{person}{Dhruv Khattar}, \bibinfo{person}{Vaibhav
  Kumar}, \bibinfo{person}{Vasudeva Varma}, {and} \bibinfo{person}{Manish
  Gupta}.} \bibinfo{year}{2018}\natexlab{a}.
\newblock \showarticletitle{HRAM: A hybrid recurrent attention machine for news
  recommendation}. In \bibinfo{booktitle}{\emph{Proceedings of the 27th ACM
  International Conference on Information and Knowledge Management}}.
  \bibinfo{pages}{1619--1622}.
\newblock


\bibitem[\protect\citeauthoryear{Khattar, Kumar, Varma, and Gupta}{Khattar
  et~al\mbox{.}}{2018b}]%
        {Weave}
\bibfield{author}{\bibinfo{person}{Dhruv Khattar}, \bibinfo{person}{Vaibhav
  Kumar}, \bibinfo{person}{Vasudeva Varma}, {and} \bibinfo{person}{Manish
  Gupta}.} \bibinfo{year}{2018}\natexlab{b}.
\newblock \showarticletitle{Weave\&rec: A word embedding based 3-d
  convolutional network for news recommendation}. In
  \bibinfo{booktitle}{\emph{Proceedings of the 27th ACM International
  Conference on Information and Knowledge Management}}.
  \bibinfo{pages}{1855--1858}.
\newblock


\bibitem[\protect\citeauthoryear{Kumar, Khattar, Gupta, Gupta, and Varma}{Kumar
  et~al\mbox{.}}{2017}]%
        {kumar2017}
\bibfield{author}{\bibinfo{person}{Vaibhav Kumar}, \bibinfo{person}{Dhruv
  Khattar}, \bibinfo{person}{Shashank Gupta}, \bibinfo{person}{Manish Gupta},
  {and} \bibinfo{person}{Vasudeva Varma}.} \bibinfo{year}{2017}\natexlab{}.
\newblock \showarticletitle{Deep Neural Architecture for News Recommendation.}.
  In \bibinfo{booktitle}{\emph{CLEF (Working Notes)}}.
\newblock


\bibitem[\protect\citeauthoryear{Li, Ren, Chen, Ren, Lian, and Ma}{Li
  et~al\mbox{.}}{2017}]%
        {NARM}
\bibfield{author}{\bibinfo{person}{Jing Li}, \bibinfo{person}{Pengjie Ren},
  \bibinfo{person}{Zhumin Chen}, \bibinfo{person}{Zhaochun Ren},
  \bibinfo{person}{Tao Lian}, {and} \bibinfo{person}{Jun Ma}.}
  \bibinfo{year}{2017}\natexlab{}.
\newblock \showarticletitle{Neural attentive session-based recommendation}.
\newblock \bibinfo{journal}{\emph{International Conference on Information and
  Knowledge Management (CIKM)}} (\bibinfo{year}{2017}).
\newblock


\bibitem[\protect\citeauthoryear{Li, Wang, and McAuley}{Li
  et~al\mbox{.}}{2020}]%
        {TISASRec}
\bibfield{author}{\bibinfo{person}{Jiacheng Li}, \bibinfo{person}{Yujie Wang},
  {and} \bibinfo{person}{Julian McAuley}.} \bibinfo{year}{2020}\natexlab{}.
\newblock \showarticletitle{Time Interval Aware Self-Attention for Sequential
  Recommendation}.
\newblock \bibinfo{journal}{\emph{Web Search and Data Mining (WSDM)}}
  (\bibinfo{year}{2020}).
\newblock


\bibitem[\protect\citeauthoryear{Li, Wang, Li, Knox, and Padmanabhan}{Li
  et~al\mbox{.}}{2011a}]%
        {SCENE}
\bibfield{author}{\bibinfo{person}{Lei Li}, \bibinfo{person}{Dingding Wang},
  \bibinfo{person}{Tao Li}, \bibinfo{person}{Daniel Knox}, {and}
  \bibinfo{person}{Balaji Padmanabhan}.} \bibinfo{year}{2011}\natexlab{a}.
\newblock \showarticletitle{Scene: a scalable two-stage personalized news
  recommendation system}. In \bibinfo{booktitle}{\emph{Proceedings of the 34th
  international ACM SIGIR conference on Research and development in Information
  Retrieval}}. \bibinfo{pages}{125--134}.
\newblock


\bibitem[\protect\citeauthoryear{Li, Zheng, and Li}{Li et~al\mbox{.}}{2011b}]%
        {logo}
\bibfield{author}{\bibinfo{person}{Lei Li}, \bibinfo{person}{Li Zheng}, {and}
  \bibinfo{person}{Tao Li}.} \bibinfo{year}{2011}\natexlab{b}.
\newblock \showarticletitle{Logo: a long-short user interest integration in
  personalized news recommendation}. In \bibinfo{booktitle}{\emph{Proceedings
  of the fifth ACM conference on Recommender systems}}.
  \bibinfo{pages}{317--320}.
\newblock


\bibitem[\protect\citeauthoryear{Liu, Dolan, and Pedersen}{Liu
  et~al\mbox{.}}{2010}]%
        {liu2010personalized}
\bibfield{author}{\bibinfo{person}{Jiahui Liu}, \bibinfo{person}{Peter Dolan},
  {and} \bibinfo{person}{Elin~R{\o}nby Pedersen}.}
  \bibinfo{year}{2010}\natexlab{}.
\newblock \showarticletitle{Personalized news recommendation based on click
  behavior}. In \bibinfo{booktitle}{\emph{Proceedings of the 15th international
  conference on Intelligent user interfaces}}. \bibinfo{pages}{31--40}.
\newblock


\bibitem[\protect\citeauthoryear{Liu, Zeng, Mokhosi, and Zhang}{Liu
  et~al\mbox{.}}{2018}]%
        {STAMP}
\bibfield{author}{\bibinfo{person}{Qiao Liu}, \bibinfo{person}{Yifu Zeng},
  \bibinfo{person}{Refuoe Mokhosi}, {and} \bibinfo{person}{Haibin Zhang}.}
  \bibinfo{year}{2018}\natexlab{}.
\newblock \showarticletitle{STAMP: short-term attention/memory priority model
  for session-based recommendation}.
\newblock \bibinfo{journal}{\emph{ACM SIGKDD International Conference on
  Knowledge Discovery \& Data Mining}} (\bibinfo{year}{2018}).
\newblock


\bibitem[\protect\citeauthoryear{Liu, Dong, and Chai}{Liu
  et~al\mbox{.}}{2016}]%
        {liu2016research}
\bibfield{author}{\bibinfo{person}{Shan Liu}, \bibinfo{person}{Yao Dong}, {and}
  \bibinfo{person}{Jianping Chai}.} \bibinfo{year}{2016}\natexlab{}.
\newblock \showarticletitle{Research of personalized news recommendation system
  based on hybrid collaborative filtering algorithm}. In
  \bibinfo{booktitle}{\emph{2016 2nd IEEE International Conference on Computer
  and Communications (ICCC)}}. IEEE, \bibinfo{pages}{865--869}.
\newblock


\bibitem[\protect\citeauthoryear{Okura, Tagami, Ono, and Tajima}{Okura
  et~al\mbox{.}}{2017}]%
        {Okura}
\bibfield{author}{\bibinfo{person}{Shumpei Okura}, \bibinfo{person}{Yukihiro
  Tagami}, \bibinfo{person}{Shingo Ono}, {and} \bibinfo{person}{Akira Tajima}.}
  \bibinfo{year}{2017}\natexlab{}.
\newblock \showarticletitle{Embedding-based news recommendation for millions of
  users}. In \bibinfo{booktitle}{\emph{Proceedings of the 23rd ACM SIGKDD
  International Conference on Knowledge Discovery and Data Mining}}.
  \bibinfo{pages}{1933--1942}.
\newblock


\bibitem[\protect\citeauthoryear{Park, Lee, and Choi}{Park
  et~al\mbox{.}}{2017}]%
        {park2017deep}
\bibfield{author}{\bibinfo{person}{Keunchan Park}, \bibinfo{person}{Jisoo Lee},
  {and} \bibinfo{person}{Jaeho Choi}.} \bibinfo{year}{2017}\natexlab{}.
\newblock \showarticletitle{Deep neural networks for news recommendations}. In
  \bibinfo{booktitle}{\emph{Proceedings of the 2017 ACM on Conference on
  Information and Knowledge Management}}. \bibinfo{pages}{2255--2258}.
\newblock


\bibitem[\protect\citeauthoryear{Paszke, Gross, Massa, Lerer, Bradbury, Chanan,
  Killeen, Lin, Gimelshein, Antiga, Desmaison, Kopf, Yang, DeVito, Raison,
  Tejani, Chilamkurthy, Steiner, Fang, Bai, and Chintala}{Paszke
  et~al\mbox{.}}{2019}]%
        {pytorch}
\bibfield{author}{\bibinfo{person}{Adam Paszke}, \bibinfo{person}{Sam Gross},
  \bibinfo{person}{Francisco Massa}, \bibinfo{person}{Adam Lerer},
  \bibinfo{person}{James Bradbury}, \bibinfo{person}{Gregory Chanan},
  \bibinfo{person}{Trevor Killeen}, \bibinfo{person}{Zeming Lin},
  \bibinfo{person}{Natalia Gimelshein}, \bibinfo{person}{Luca Antiga},
  \bibinfo{person}{Alban Desmaison}, \bibinfo{person}{Andreas Kopf},
  \bibinfo{person}{Edward Yang}, \bibinfo{person}{Zachary DeVito},
  \bibinfo{person}{Martin Raison}, \bibinfo{person}{Alykhan Tejani},
  \bibinfo{person}{Sasank Chilamkurthy}, \bibinfo{person}{Benoit Steiner},
  \bibinfo{person}{Lu Fang}, \bibinfo{person}{Junjie Bai}, {and}
  \bibinfo{person}{Soumith Chintala}.} \bibinfo{year}{2019}\natexlab{}.
\newblock \showarticletitle{PyTorch: An Imperative Style, High-Performance Deep
  Learning Library}.
\newblock In \bibinfo{booktitle}{\emph{Advances in Neural Information
  Processing Systems 32}}.
\newblock


\bibitem[\protect\citeauthoryear{Phelan, McCarthy, Bennett, and Smyth}{Phelan
  et~al\mbox{.}}{2011}]%
        {phelan2011terms}
\bibfield{author}{\bibinfo{person}{Owen Phelan}, \bibinfo{person}{Kevin
  McCarthy}, \bibinfo{person}{Mike Bennett}, {and} \bibinfo{person}{Barry
  Smyth}.} \bibinfo{year}{2011}\natexlab{}.
\newblock \showarticletitle{Terms of a feather: Content-based news
  recommendation and discovery using twitter}. In
  \bibinfo{booktitle}{\emph{European Conference on Information Retrieval}}.
  Springer, \bibinfo{pages}{448--459}.
\newblock


\bibitem[\protect\citeauthoryear{Phelan, McCarthy, and Smyth}{Phelan
  et~al\mbox{.}}{2009}]%
        {phelan2009using}
\bibfield{author}{\bibinfo{person}{Owen Phelan}, \bibinfo{person}{Kevin
  McCarthy}, {and} \bibinfo{person}{Barry Smyth}.}
  \bibinfo{year}{2009}\natexlab{}.
\newblock \showarticletitle{Using twitter to recommend real-time topical news}.
  In \bibinfo{booktitle}{\emph{Proceedings of the third ACM conference on
  Recommender systems}}. \bibinfo{pages}{385--388}.
\newblock


\bibitem[\protect\citeauthoryear{Qi, Wu, Wu, and Huang}{Qi
  et~al\mbox{.}}{2021}]%
        {pprec}
\bibfield{author}{\bibinfo{person}{Tao Qi}, \bibinfo{person}{Fangzhao Wu},
  \bibinfo{person}{Chuhan Wu}, {and} \bibinfo{person}{Yongfeng Huang}.}
  \bibinfo{year}{2021}\natexlab{}.
\newblock \showarticletitle{PP-Rec: News Recommendation with Personalized User
  Interest and Time-aware News Popularity}.
\newblock \bibinfo{journal}{\emph{arXiv preprint arXiv:2106.01300}}
  (\bibinfo{year}{2021}).
\newblock


\bibitem[\protect\citeauthoryear{Quadrana, Karatzoglou, Hidasi, and
  Cremonesi}{Quadrana et~al\mbox{.}}{2017}]%
        {HIERARCHICALGRU4REC}
\bibfield{author}{\bibinfo{person}{Massimo Quadrana},
  \bibinfo{person}{Alexandros Karatzoglou}, \bibinfo{person}{Bal{\'a}zs
  Hidasi}, {and} \bibinfo{person}{Paolo Cremonesi}.}
  \bibinfo{year}{2017}\natexlab{}.
\newblock \showarticletitle{Personalizing session-based recommendations with
  hierarchical recurrent neural networks}.
\newblock \bibinfo{journal}{\emph{Recommender Systems (RecSys)}}
  (\bibinfo{year}{2017}).
\newblock


\bibitem[\protect\citeauthoryear{Schifferer, Titericz, Deotte, Henkel, Onodera,
  Liu, Tunguz, Oldridge, De~Souza Pereira~Moreira, and Erdem}{Schifferer
  et~al\mbox{.}}{2020}]%
        {recsys2020challengewinner}
\bibfield{author}{\bibinfo{person}{Benedikt Schifferer},
  \bibinfo{person}{Gilberto Titericz}, \bibinfo{person}{Chris Deotte},
  \bibinfo{person}{Christof Henkel}, \bibinfo{person}{Kazuki Onodera},
  \bibinfo{person}{Jiwei Liu}, \bibinfo{person}{Bojan Tunguz},
  \bibinfo{person}{Even Oldridge}, \bibinfo{person}{Gabriel De~Souza
  Pereira~Moreira}, {and} \bibinfo{person}{Ahmet Erdem}.}
  \bibinfo{year}{2020}\natexlab{}.
\newblock \showarticletitle{GPU Accelerated Feature Engineering and Training
  for Recommender Systems}.
\newblock In \bibinfo{booktitle}{\emph{Proceedings of the Recommender Systems
  Challenge 2020}}. \bibinfo{pages}{16--23}.
\newblock


\bibitem[\protect\citeauthoryear{Son, Kim, and Park}{Son et~al\mbox{.}}{2013}]%
        {son2013location}
\bibfield{author}{\bibinfo{person}{Jeong-Woo Son}, \bibinfo{person}{A-Yeong
  Kim}, {and} \bibinfo{person}{Seong-Bae Park}.}
  \bibinfo{year}{2013}\natexlab{}.
\newblock \showarticletitle{A location-based news article recommendation with
  explicit localized semantic analysis}. In
  \bibinfo{booktitle}{\emph{Proceedings of the 36th international ACM SIGIR
  conference on Research and development in information retrieval}}.
  \bibinfo{pages}{293--302}.
\newblock


\bibitem[\protect\citeauthoryear{Song, Elkahky, and He}{Song
  et~al\mbox{.}}{2016}]%
        {song2016}
\bibfield{author}{\bibinfo{person}{Yang Song}, \bibinfo{person}{Ali~Mamdouh
  Elkahky}, {and} \bibinfo{person}{Xiaodong He}.}
  \bibinfo{year}{2016}\natexlab{}.
\newblock \showarticletitle{Multi-rate deep learning for temporal
  recommendation}. In \bibinfo{booktitle}{\emph{Proceedings of the 39th
  International ACM SIGIR conference on Research and Development in Information
  Retrieval}}. \bibinfo{pages}{909--912}.
\newblock


\bibitem[\protect\citeauthoryear{Sun, Liu, Wu, Pei, Lin, Ou, and Jiang}{Sun
  et~al\mbox{.}}{2019}]%
        {BERT4Rec}
\bibfield{author}{\bibinfo{person}{Fei Sun}, \bibinfo{person}{Jun Liu},
  \bibinfo{person}{Jian Wu}, \bibinfo{person}{Changhua Pei},
  \bibinfo{person}{Xiao Lin}, \bibinfo{person}{Wenwu Ou}, {and}
  \bibinfo{person}{Peng Jiang}.} \bibinfo{year}{2019}\natexlab{}.
\newblock \showarticletitle{BERT4Rec: Sequential recommendation with
  bidirectional encoder representations from transformer}.
\newblock \bibinfo{journal}{\emph{International Conference on Information and
  Knowledge Management (CIKM)}} (\bibinfo{year}{2019}).
\newblock


\bibitem[\protect\citeauthoryear{Tavakolifard, Gulla, Almeroth, Ingvaldesn,
  Nygreen, and Berg}{Tavakolifard et~al\mbox{.}}{2013}]%
        {tailored}
\bibfield{author}{\bibinfo{person}{Mozhgan Tavakolifard},
  \bibinfo{person}{Jon~Atle Gulla}, \bibinfo{person}{Kevin~C Almeroth},
  \bibinfo{person}{Jon~Espen Ingvaldesn}, \bibinfo{person}{Gaute Nygreen},
  {and} \bibinfo{person}{Erik Berg}.} \bibinfo{year}{2013}\natexlab{}.
\newblock \showarticletitle{Tailored news in the palm of your hand: a
  multi-perspective transparent approach to news recommendation}. In
  \bibinfo{booktitle}{\emph{Proceedings of the 22nd international conference on
  world wide web}}. \bibinfo{pages}{305--308}.
\newblock


\bibitem[\protect\citeauthoryear{Vaswani, Shazeer, Parmar, Uszkoreit, Jones,
  Gomez, Kaiser, and Polosukhin}{Vaswani et~al\mbox{.}}{2017}]%
        {TRANSFORMER}
\bibfield{author}{\bibinfo{person}{Ashish Vaswani}, \bibinfo{person}{Noam
  Shazeer}, \bibinfo{person}{Niki Parmar}, \bibinfo{person}{Jakob Uszkoreit},
  \bibinfo{person}{Llion Jones}, \bibinfo{person}{Aidan~N Gomez},
  \bibinfo{person}{{\L}ukasz Kaiser}, {and} \bibinfo{person}{Illia
  Polosukhin}.} \bibinfo{year}{2017}\natexlab{}.
\newblock \showarticletitle{Attention is all you need}.
\newblock \bibinfo{journal}{\emph{Advances in neural information processing
  systems}} (\bibinfo{year}{2017}).
\newblock


\bibitem[\protect\citeauthoryear{Viana and Soares}{Viana and Soares}{2017}]%
        {viana2017hybrid}
\bibfield{author}{\bibinfo{person}{Paula Viana} {and}
  \bibinfo{person}{M{\'a}rcio Soares}.} \bibinfo{year}{2017}\natexlab{}.
\newblock \showarticletitle{A hybrid approach for personalized news
  recommendation in a mobility scenario using long-short user interest}.
\newblock \bibinfo{journal}{\emph{International Journal on Artificial
  Intelligence Tools}} \bibinfo{volume}{26}, \bibinfo{number}{02}
  (\bibinfo{year}{2017}), \bibinfo{pages}{1760012}.
\newblock


\bibitem[\protect\citeauthoryear{Wang, Wu, Liu, and Xie}{Wang
  et~al\mbox{.}}{2020}]%
        {FIM}
\bibfield{author}{\bibinfo{person}{Heyuan Wang}, \bibinfo{person}{Fangzhao Wu},
  \bibinfo{person}{Zheng Liu}, {and} \bibinfo{person}{Xing Xie}.}
  \bibinfo{year}{2020}\natexlab{}.
\newblock \showarticletitle{Fine-grained Interest Matching for Neural News
  Recommendation}. In \bibinfo{booktitle}{\emph{Proceedings of the 58th Annual
  Meeting of the Association for Computational Linguistics}}.
  \bibinfo{pages}{836--845}.
\newblock


\bibitem[\protect\citeauthoryear{Wang, Zhang, Xie, and Guo}{Wang
  et~al\mbox{.}}{2018}]%
        {DKNwang2018a}
\bibfield{author}{\bibinfo{person}{Hongwei Wang}, \bibinfo{person}{Fuzheng
  Zhang}, \bibinfo{person}{Xing Xie}, {and} \bibinfo{person}{Minyi Guo}.}
  \bibinfo{year}{2018}\natexlab{}.
\newblock \showarticletitle{DKN: Deep knowledge-aware network for news
  recommendation}. In \bibinfo{booktitle}{\emph{Proceedings of the 2018 world
  wide web conference}}. \bibinfo{pages}{1835--1844}.
\newblock


\bibitem[\protect\citeauthoryear{Wei, Feng, Chen, Wu, Yi, and He}{Wei
  et~al\mbox{.}}{2021}]%
        {modelagnosticcounterfactual}
\bibfield{author}{\bibinfo{person}{Tianxin Wei}, \bibinfo{person}{Fuli Feng},
  \bibinfo{person}{Jiawei Chen}, \bibinfo{person}{Ziwei Wu},
  \bibinfo{person}{Jinfeng Yi}, {and} \bibinfo{person}{Xiangnan He}.}
  \bibinfo{year}{2021}\natexlab{}.
\newblock \showarticletitle{Model-agnostic counterfactual reasoning for
  eliminating popularity bias in recommender system}. In
  \bibinfo{booktitle}{\emph{Proceedings of the 27th ACM SIGKDD Conference on
  Knowledge Discovery \& Data Mining}}. \bibinfo{pages}{1791--1800}.
\newblock


\bibitem[\protect\citeauthoryear{Wu, Cheng, Zhang, Huang, Li, and Mei}{Wu
  et~al\mbox{.}}{2017b}]%
        {wu2017sequential}
\bibfield{author}{\bibinfo{person}{Bo Wu}, \bibinfo{person}{Wen{-}Huang Cheng},
  \bibinfo{person}{Yongdong Zhang}, \bibinfo{person}{Qiushi Huang},
  \bibinfo{person}{Jintao Li}, {and} \bibinfo{person}{Tao Mei}.}
  \bibinfo{year}{2017}\natexlab{b}.
\newblock \showarticletitle{Sequential Prediction of Social Media Popularity
  with Deep Temporal Context Networks}.
\newblock \bibinfo{journal}{\emph{International Joint Conference on Artificial
  Intelligence (IJCAI)}} (\bibinfo{year}{2017}).
\newblock


\bibitem[\protect\citeauthoryear{Wu, Wu, An, Huang, Huang, and Xie}{Wu
  et~al\mbox{.}}{2019a}]%
        {NAML}
\bibfield{author}{\bibinfo{person}{Chuhan Wu}, \bibinfo{person}{Fangzhao Wu},
  \bibinfo{person}{Mingxiao An}, \bibinfo{person}{Jianqiang Huang},
  \bibinfo{person}{Yongfeng Huang}, {and} \bibinfo{person}{Xing Xie}.}
  \bibinfo{year}{2019}\natexlab{a}.
\newblock \showarticletitle{Neural news recommendation with attentive
  multi-view learning}.
\newblock \bibinfo{journal}{\emph{arXiv preprint arXiv:1907.05576}}
  (\bibinfo{year}{2019}).
\newblock


\bibitem[\protect\citeauthoryear{Wu, Wu, An, Huang, Huang, and Xie}{Wu
  et~al\mbox{.}}{2019b}]%
        {NPA}
\bibfield{author}{\bibinfo{person}{Chuhan Wu}, \bibinfo{person}{Fangzhao Wu},
  \bibinfo{person}{Mingxiao An}, \bibinfo{person}{Jianqiang Huang},
  \bibinfo{person}{Yongfeng Huang}, {and} \bibinfo{person}{Xing Xie}.}
  \bibinfo{year}{2019}\natexlab{b}.
\newblock \showarticletitle{NPA: neural news recommendation with personalized
  attention}. In \bibinfo{booktitle}{\emph{Proceedings of the 25th ACM SIGKDD
  International Conference on Knowledge Discovery \& Data Mining}}.
  \bibinfo{pages}{2576--2584}.
\newblock


\bibitem[\protect\citeauthoryear{Wu, Wu, Ge, Qi, Huang, and Xie}{Wu
  et~al\mbox{.}}{2019c}]%
        {NRMS}
\bibfield{author}{\bibinfo{person}{Chuhan Wu}, \bibinfo{person}{Fangzhao Wu},
  \bibinfo{person}{Suyu Ge}, \bibinfo{person}{Tao Qi},
  \bibinfo{person}{Yongfeng Huang}, {and} \bibinfo{person}{Xing Xie}.}
  \bibinfo{year}{2019}\natexlab{c}.
\newblock \showarticletitle{Neural news recommendation with multi-head
  self-attention}. In \bibinfo{booktitle}{\emph{Proceedings of the 2019
  Conference on Empirical Methods in Natural Language Processing and the 9th
  International Joint Conference on Natural Language Processing
  (EMNLP-IJCNLP)}}. \bibinfo{pages}{6390--6395}.
\newblock


\bibitem[\protect\citeauthoryear{Wu, Wu, and Huang}{Wu et~al\mbox{.}}{2021a}]%
        {wuchuhansurvey}
\bibfield{author}{\bibinfo{person}{Chuhan Wu}, \bibinfo{person}{Fangzhao Wu},
  {and} \bibinfo{person}{Yongfeng Huang}.} \bibinfo{year}{2021}\natexlab{a}.
\newblock \showarticletitle{Personalized News Recommendation: A Survey}.
\newblock \bibinfo{journal}{\emph{arXiv preprint arXiv:2106.08934}}
  (\bibinfo{year}{2021}).
\newblock


\bibitem[\protect\citeauthoryear{Wu, Wu, Qi, and Huang}{Wu
  et~al\mbox{.}}{2021b}]%
        {wu2021empowering}
\bibfield{author}{\bibinfo{person}{Chuhan Wu}, \bibinfo{person}{Fangzhao Wu},
  \bibinfo{person}{Tao Qi}, {and} \bibinfo{person}{Yongfeng Huang}.}
  \bibinfo{year}{2021}\natexlab{b}.
\newblock \showarticletitle{Empowering News Recommendation with Pre-trained
  Language Models}.
\newblock \bibinfo{journal}{\emph{arXiv preprint arXiv:2104.07413}}
  (\bibinfo{year}{2021}).
\newblock


\bibitem[\protect\citeauthoryear{Wu, Wu, Qi, and Huang}{Wu
  et~al\mbox{.}}{2021c}]%
        {wu2021feedrec}
\bibfield{author}{\bibinfo{person}{Chuhan Wu}, \bibinfo{person}{Fangzhao Wu},
  \bibinfo{person}{Tao Qi}, {and} \bibinfo{person}{Yongfeng Huang}.}
  \bibinfo{year}{2021}\natexlab{c}.
\newblock \showarticletitle{FeedRec: News Feed Recommendation with Various User
  Feedbacks}.
\newblock \bibinfo{journal}{\emph{arXiv preprint arXiv:2102.04903}}
  (\bibinfo{year}{2021}).
\newblock


\bibitem[\protect\citeauthoryear{Wu, Wu, Qi, and Huang}{Wu
  et~al\mbox{.}}{2021d}]%
        {MMRec}
\bibfield{author}{\bibinfo{person}{Chuhan Wu}, \bibinfo{person}{Fangzhao Wu},
  \bibinfo{person}{Tao Qi}, {and} \bibinfo{person}{Yongfeng Huang}.}
  \bibinfo{year}{2021}\natexlab{d}.
\newblock \showarticletitle{MM-Rec: Multimodal News Recommendation}.
\newblock \bibinfo{journal}{\emph{arXiv preprint arXiv:2104.07407}}
  (\bibinfo{year}{2021}).
\newblock


\bibitem[\protect\citeauthoryear{Wu, Ahmed, Beutel, Smola, and Jing}{Wu
  et~al\mbox{.}}{2017a}]%
        {wu2017recurrent}
\bibfield{author}{\bibinfo{person}{Chao-Yuan Wu}, \bibinfo{person}{Amr Ahmed},
  \bibinfo{person}{Alex Beutel}, \bibinfo{person}{Alexander~J Smola}, {and}
  \bibinfo{person}{How Jing}.} \bibinfo{year}{2017}\natexlab{a}.
\newblock \showarticletitle{Recurrent recommender networks}.
\newblock \bibinfo{journal}{\emph{Web Search and Data Mining (WSDM)}}
  (\bibinfo{year}{2017}).
\newblock


\bibitem[\protect\citeauthoryear{Wu, Qiao, Chen, Wu, Qi, Lian, Liu, Xie, Gao,
  Wu, et~al\mbox{.}}{Wu et~al\mbox{.}}{2020}]%
        {minddataset}
\bibfield{author}{\bibinfo{person}{Fangzhao Wu}, \bibinfo{person}{Ying Qiao},
  \bibinfo{person}{Jiun-Hung Chen}, \bibinfo{person}{Chuhan Wu},
  \bibinfo{person}{Tao Qi}, \bibinfo{person}{Jianxun Lian},
  \bibinfo{person}{Danyang Liu}, \bibinfo{person}{Xing Xie},
  \bibinfo{person}{Jianfeng Gao}, \bibinfo{person}{Winnie Wu}, {et~al\mbox{.}}}
  \bibinfo{year}{2020}\natexlab{}.
\newblock \showarticletitle{Mind: A large-scale dataset for news
  recommendation}. In \bibinfo{booktitle}{\emph{Proceedings of the 58th Annual
  Meeting of the Association for Computational Linguistics}}.
  \bibinfo{pages}{3597--3606}.
\newblock


\bibitem[\protect\citeauthoryear{Yang, Dai, Yang, Carbonell, Salakhutdinov, and
  Le}{Yang et~al\mbox{.}}{2019}]%
        {XLNET}
\bibfield{author}{\bibinfo{person}{Zhilin Yang}, \bibinfo{person}{Zihang Dai},
  \bibinfo{person}{Yiming Yang}, \bibinfo{person}{Jaime Carbonell},
  \bibinfo{person}{Russ~R Salakhutdinov}, {and} \bibinfo{person}{Quoc~V Le}.}
  \bibinfo{year}{2019}\natexlab{}.
\newblock \showarticletitle{Xlnet: Generalized autoregressive pretraining for
  language understanding}.
\newblock \bibinfo{journal}{\emph{Advances in Neural Information Processing
  Systems (NeurIPS)}} (\bibinfo{year}{2019}).
\newblock


\bibitem[\protect\citeauthoryear{Yeung and Yang}{Yeung and Yang}{2010}]%
        {yeung2010proactive}
\bibfield{author}{\bibinfo{person}{Kam~Fung Yeung} {and}
  \bibinfo{person}{Yanyan Yang}.} \bibinfo{year}{2010}\natexlab{}.
\newblock \showarticletitle{A proactive personalized mobile news recommendation
  system}. In \bibinfo{booktitle}{\emph{2010 Developments in E-systems
  Engineering}}. IEEE, \bibinfo{pages}{207--212}.
\newblock


\bibitem[\protect\citeauthoryear{Yi and Chang}{Yi and Chang}{2021}]%
        {hyperconnectboosting}
\bibfield{author}{\bibinfo{person}{Joonyoung Yi} {and} \bibinfo{person}{Buru
  Chang}.} \bibinfo{year}{2021}\natexlab{}.
\newblock \showarticletitle{Efficient Click-Through Rate Prediction for
  Developing Countries via Tabular Learning}.
\newblock \bibinfo{journal}{\emph{arXiv preprint arXiv:2104.07553}}
  (\bibinfo{year}{2021}).
\newblock


\bibitem[\protect\citeauthoryear{Yu, Liu, Wu, Wang, and Tan}{Yu
  et~al\mbox{.}}{2016}]%
        {DREAM}
\bibfield{author}{\bibinfo{person}{Feng Yu}, \bibinfo{person}{Qiang Liu},
  \bibinfo{person}{Shu Wu}, \bibinfo{person}{Liang Wang}, {and}
  \bibinfo{person}{Tieniu Tan}.} \bibinfo{year}{2016}\natexlab{}.
\newblock \showarticletitle{A dynamic recurrent model for next basket
  recommendation}.
\newblock \bibinfo{journal}{\emph{ACM SIGIR conference on Research and
  Development in Information Retrieval}} (\bibinfo{year}{2016}).
\newblock


\bibitem[\protect\citeauthoryear{Zhang, Liu, and Gulla}{Zhang
  et~al\mbox{.}}{2019}]%
        {zhang2019a}
\bibfield{author}{\bibinfo{person}{Lemei Zhang}, \bibinfo{person}{Peng Liu},
  {and} \bibinfo{person}{Jon~Atle Gulla}.} \bibinfo{year}{2019}\natexlab{}.
\newblock \showarticletitle{Dynamic attention-integrated neural network for
  session-based news recommendation}.
\newblock \bibinfo{journal}{\emph{Machine Learning}} \bibinfo{volume}{108},
  \bibinfo{number}{10} (\bibinfo{year}{2019}), \bibinfo{pages}{1851--1875}.
\newblock


\bibitem[\protect\citeauthoryear{Zhang, Dai, Xu, Feng, Wang, Bian, Wang, and
  Liu}{Zhang et~al\mbox{.}}{2014}]%
        {RNN4Rec}
\bibfield{author}{\bibinfo{person}{Yuyu Zhang}, \bibinfo{person}{Hanjun Dai},
  \bibinfo{person}{Chang Xu}, \bibinfo{person}{Jun Feng},
  \bibinfo{person}{Taifeng Wang}, \bibinfo{person}{Jiang Bian},
  \bibinfo{person}{Bin Wang}, {and} \bibinfo{person}{Tie-Yan Liu}.}
  \bibinfo{year}{2014}\natexlab{}.
\newblock \showarticletitle{Sequential click prediction for sponsored search
  with recurrent neural networks}.
\newblock \bibinfo{journal}{\emph{AAAI Conference on Artificial Intelligence
  (AAAI)}} (\bibinfo{year}{2014}).
\newblock


\bibitem[\protect\citeauthoryear{Zhou, Mou, Fan, Pi, Bian, Zhou, Zhu, and
  Gai}{Zhou et~al\mbox{.}}{2019}]%
        {DINEVOLUTION}
\bibfield{author}{\bibinfo{person}{Guorui Zhou}, \bibinfo{person}{Na Mou},
  \bibinfo{person}{Ying Fan}, \bibinfo{person}{Qi Pi}, \bibinfo{person}{Weijie
  Bian}, \bibinfo{person}{Chang Zhou}, \bibinfo{person}{Xiaoqiang Zhu}, {and}
  \bibinfo{person}{Kun Gai}.} \bibinfo{year}{2019}\natexlab{}.
\newblock \showarticletitle{Deep interest evolution network for click-through
  rate prediction}.
\newblock \bibinfo{journal}{\emph{AAAI Conference on Artificial Intelligence
  (AAAI)}} (\bibinfo{year}{2019}).
\newblock


\bibitem[\protect\citeauthoryear{Zhu, Zhou, Song, Tan, and Guo}{Zhu
  et~al\mbox{.}}{2019}]%
        {zhu2019}
\bibfield{author}{\bibinfo{person}{Qiannan Zhu}, \bibinfo{person}{Xiaofei
  Zhou}, \bibinfo{person}{Zeliang Song}, \bibinfo{person}{Jianlong Tan}, {and}
  \bibinfo{person}{Li Guo}.} \bibinfo{year}{2019}\natexlab{}.
\newblock \showarticletitle{Dan: Deep attention neural network for news
  recommendation}. In \bibinfo{booktitle}{\emph{Proceedings of the AAAI
  Conference on Artificial Intelligence}}, Vol.~\bibinfo{volume}{33}.
  \bibinfo{pages}{5973--5980}.
\newblock


\bibitem[\protect\citeauthoryear{Zihayat, Ayanso, Zhao, Davoudi, and
  An}{Zihayat et~al\mbox{.}}{2019}]%
        {utilitybased}
\bibfield{author}{\bibinfo{person}{Morteza Zihayat}, \bibinfo{person}{Anteneh
  Ayanso}, \bibinfo{person}{Xing Zhao}, \bibinfo{person}{Heidar Davoudi}, {and}
  \bibinfo{person}{Aijun An}.} \bibinfo{year}{2019}\natexlab{}.
\newblock \showarticletitle{A utility-based news recommendation system}.
\newblock \bibinfo{journal}{\emph{Decision Support Systems}}
  \bibinfo{volume}{117} (\bibinfo{year}{2019}), \bibinfo{pages}{14--27}.
\newblock


\end{thebibliography}
